\documentclass[referee,twocolumn]{aastex7}

\RequirePackage{xspace}
\RequirePackage{amsmath}
\RequirePackage{amsfonts}
\RequirePackage{amssymb}

\def\ifundefined#1{\expandafter\ifx\csname#1\endcsname\relax}

\makeatletter
\newcommand*{\rom}[1]{\expandafter\@slowromancap\romannumeral #1@}
\makeatother

\def\la{\mathrel{\hbox{\rlap{\hbox{\lower4pt\hbox{$\sim$}}}\hbox{$<$}}}}
\def\ga{\mathrel{\hbox{\rlap{\hbox{\lower4pt\hbox{$\sim$}}}\hbox{$>$}}}}

\newcommand{\be}{\begin{equation}}
\newcommand{\ee}{\end{equation}}

\newcommand{\bea}{\begin{eqnarray}}
\newcommand{\eea}{\end{eqnarray}}

\ifundefined{ensuremath}\def\ensuremath#1{\relax\ifmmode{#1}}
\else${#1}$\fi\else\relax\fi
\ifundefined{nuc}\def\nuc#1#2{\relax\ifmmode{}^{#1}{\protect\text{#2}}
\else${}^{#1}$#2\fi}\else\relax\fi

\ifundefined{ion}
\newcommand\ion[2]{#1$\;${%
\ifx\@currsize\normalsize\small \else
\ifx\@currsize\small\footnotesize \else
\ifx\@currsize\footnotesize\scriptsize \else
\ifx\@currsize\scriptsize\tiny \else
\ifx\@currsize\large\normalsize \else
\ifx\@currsize\Large\large
\fi\fi\fi\fi\fi\fi
\rmfamily\rom{#2}}\relax}%
\else\relax\fi

\newcommand{\kmps}{\ensuremath{\text{km}~\text{s}^{-1}}\xspace}

\newcommand{\msol}{\ensuremath{{\text{M}_\odot}}\xspace}

\newcommand{\phx}{\texttt{PHOENIX}\xspace}
\newcommand\phxO{\texttt{PHOENIX/1D}\xspace}

\newcommand{\sneia}{SNe~I\lowercase{a}\xspace}

\makeatletter

\newcommand\Autoref[1]{\@first@ref#1,@}
\def\@throw@dot#1.#2@{#1}% discard everything after the dot
\def\@set@refname#1{%    % set \@refname to autoefname+s using \getrefbykeydefault
    \edef\@tmp{\getrefbykeydefault{#1}{anchor}{}}%
    \def\@refname{\@nameuse{\expandafter\@throw@dot\@tmp.@autorefname}s}%
}
\def\@first@ref#1,#2{%
  \ifx#2@\autoref{#1}\let\@nextref\@gobble% only one ref, revert to normal \autoref
  \else%
    \@set@refname{#1}%  set \@refname to autoref name
    \@refname~\ref{#1}% add autoefname and first reference
    \let\@nextref\@next@ref% push processing to \@next@ref
  \fi%
  \@nextref#2%
}
\def\@next@ref#1,#2{%
   \ifx#2@ and~\ref{#1}\let\@nextref\@gobble% at end: print and+\ref and stop
   \else, \ref{#1}% print  ,+\ref and continue
   \fi%
   \@nextref#2%
}

\makeatother
\usepackage{new_aas_macros}

\xspaceaddexceptions{]}

\newcommand{\microns}{$\mu$m\xspace}

\newcommand{\Msun}{\ensuremath{\text{M}_{\odot}}\xspace}

\newcommand{\vjm}{SN~2024vjm\xspace}

\newcommand{\jwst}{\textit{JWST}\xspace}

\newcommand{\texp}{\ensuremath{\text{t}_{exp}}\xspace}
\newcommand{\tbmax}{\ensuremath{\text{t}_{{B}_{max}}}\xspace}
\newcommand{\imax}{\ensuremath{i_\text{max}}\xspace}
\newcommand{\Nifs}{$^{56}$Ni\xspace}
\newcommand{\pxl}{SN~2024pxl\xspace}

\newcommand{\sniax}{SN~Iax\xspace}
\newcommand{\sneiax}{SNe~Iax\xspace}
\newcommand{\phasebmax}{196.6\xspace}

\newcommand{\phasebmaxrf}{196.1\xspace}
\newcommand{\phaseexprf}{202.8\xspace}
\newcommand{\wsiax}{W7-Iax\xspace}

\makeatletter
\def\parse#1#2#3#4#5{\@parse#4\@nil}

\def\@parse#1.#2\@nil{
  \def\current@type{#1}
  \ifdefstring{\current@type}{figure}{\@@parse#2\@nil}
    {\ifdefstring{\current@type}{table}{\@@parse#2\@nil}}
    {}
  }
\def\@@parse#1.#2\@nil{\def\current@number{#2}}

\newcounter{highestfigureyet}
\newcounter{highesttableyet}
\def\checkfloatreforder#1{  \ifcsname r@#1\endcsname     \expandafter\expandafter\expandafter\parse         \csname r@#1\endcsname  \else\def\current@type{0}\def\current@number{0}\fi  \ifdefstring{\current@type}{figure}{    \ifnumgreater{\current@number}{\value{highestfigureyet}}
      {\setcounter{highestfigureyet}{\current@number}}
      {\ifnumgreater{\value{highestfigureyet}}{\current@number}
        {\expandafter\providebool{figure@\current@number @isrefd}         \expandafter\ifbool{figure@\current@number @isrefd}
            {}{\GenericWarning{}{Warning! Figure \current@number\space referenced after
     figure \thehighestfigureyet\space on page \thepage\space without being referenced before}}}{}}    \expandafter\providebool{figure@\current@number @isrefd}    \expandafter\booltrue{figure@\current@number @isrefd}  }{    \ifdefstring{\current@type}{table}{      \ifnumgreater{\current@number}{\value{highesttableyet}}
        {\setcounter{highesttableyet}{\current@number}}
        {\ifnumgreater{\value{highesttableyet}}{\current@number}
          {\expandafter\providebool{table@\current@number @isrefd}           \expandafter\ifbool{table@\current@number @isrefd}
              {}{\GenericWarning{}{Warning! Table \current@number\space referenced after
     table \thehighesttableyet\space on page \thepage\space without being referenced before}}}{}}      \expandafter\providebool{table@\current@number @isrefd}      \expandafter\booltrue{table@\current@number @isrefd}      }{}  
  }}

\ifundefined{fref}
  \checkfloatreforder{#1}
\else
\let\old@fref\fref
\renewcommand*{\fref}[1]{\old@fref{#1}
  \checkfloatreforder{#1}}
\fi

\makeatother

\graphicspath{{./figs/}}

\begin{document}

\title{Panchromatic JWST Observations and Models of the Dim Type Iax
  Supernova \vjm at 200 days.}

\newcommand{\PSI}{\affiliation{Planetary Science Institute, 1700 East Fort
  Lowell Road, Suite 106,Tucson, AZ 85719-2395 USA}}
\newcommand{\HS}{\affiliation{Hamburger Sternwarte, Gojenbergsweg 112, 21029 Hamburg, Germany}}
\newcommand{\IFA}{\affiliation{Institute for Astronomy, University of Hawai’i at Manoa, 2680 Woodlawn Dr., Hawai’i, HI 96822, USA}}
\newcommand{\VT}{\affiliation{Department of Physics, Virginia Tech,
    850 West Campus  Drive, Blacksburg VA, 24061, USA}}
\newcommand{\GRFP}{\affiliation{National Science Foundation Graduate Research Fellow}}

\newcommand{\STSci}{\affiliation{Space Telescope Science Institute, 3700 San Martin Drive, Baltimore, MD 21218-2410, USA}}
\newcommand{\FSU}{\affiliation{Department of Physics, Florida State
    University, Tallahassee, FL 32306, USA}}
\newcommand{\Carnegie}{\affiliation{Observatories of the Carnegie
    Institution for Science, 813 Santa Barbara St., Pasadena, CA 91101, USA}}
\newcommand{\MSU}{\affiliation{Department of Physics \& Astronomy,
    Michigan State University, East Lansing, MI, USA}}
\newcommand{\TAMU}{\affiliation{George P. and Cynthia Woods Mitchell
    Institute for Fundamental Physics and Astronomy,
    Department of Physics and Astronomy, Texas 
             A\&M University, College Station, TX 77843, USA}}
\newcommand{\IALP}{\affiliation{Instituto de Astrof\'isica de La Plata
    (IALP), CONICET, Paseo del Bosque S/N, B1900FWA La Plata, Argentina}}
\newcommand{\LaPlata}{\affiliation{Facultad de Ciencias Astron\'omicas
    y Geof\'isicas Universidad Nacional de La Plata, Paseo del Bosque,
    B1900FWA, La Plata, Argentina}}
\newcommand{\WPI}{\affiliation{Kavli Institute for the Physics and
    Mathematics of the Universe (WPI), The University of Tokyo,
    Kashiwa, 277-8583 Chiba, Japan}} 

\newcommand{\ICE}{\affiliation{Institute of Space Sciences (ICE,
    CSIC), Campus UAB, Carrer de Can Magrans, s/n, E-08193 Barcelona, Spain}}

\newcommand{\IEEC}{\affiliation{Institut d’Estudis Espacials de
    Catalunya (IEEC), E-08034  Barcelona, Spain}} 

\newcommand{\LCO}{\affiliation{Las Campanas Observatory, Carnegie
    Observatories, Casilla 601, La Serena, Chile}} 

\newcommand{\Aarhus}{\affiliation{Department of Physics and Astronomy,
    Aarhus University, Ny  Munkegade 120, DK-8000 Aarhus C, Denmark.}} 

\newcommand{\OU}{\affiliation{Homer L.~Dodge Department of Physics and
  Astronomy, University of Oklahoma, 440 W. Brooks, Rm 100, Norman, OK
  73019-2061}}  

\newcommand{\UCSC}{\affiliation{Department of Astronomy and Astrophysics,
  University of California, Santa Cruz, CA 95064, USA}} 
\newcommand{\Melbourne}{\affiliation{School of Physics, The University of
  Melbourne, VIC 3010, Australia}}

\newcommand{\LPNHE}{\affiliation{LPNHE, (CNRS/IN2P3), Sorbonne
  Universit\'e, Universit\'e Paris Cit\'e), Laboratoire de Physique
  Nucl\'eaire et de Hautes \'Energies, 75005, Paris, France}}

\newcommand{\Princeton}{\affiliation{Princeton University, 4 Ivy Lane,
    Princeton, NJ 08544, USA}}

\newcommand{\Berkeley}{\affiliation{Department of Astronomy,
    University of California, Berkeley, CA 94720-3411, USA}}

\newcommand{\Tsinghua}{\affiliation{Physics Department, Tsinghua
    University, Beijing, 100084, China}}

\newcommand{\Thailand}{\affiliation{National Astronomical Research
    Institute of Thailand, 260 Moo 4, Donkaew, Maerim, Chiang Mai,
    50180, Thailand}}

\newcommand{\UVA}{\affiliation{Department of Astronomy, University of
    Virginia, 530 McCormick Rd, Charlottesville, VA 22904, USA}}

\newcommand{\LJMU}{\affiliation{Astrophysics Research Institute,
    Liverpool John Moores University, 146 Brownlow Hill, Liverpool L3
    5RF, UK}}

\newcommand{\MPIA}{\affiliation{Max-Planck-Institut f\"ur Astrophysik,
    Karl-Schwarzschild Stra{\ss}e 1, 85748 Garching, Germany}}

\newcommand{\JHU}{\affiliation{Physics and Astronomy Department,
    Johns Hopkins University, Baltimore, MD 21218, USA}}

\newcommand{\OSU}{\affiliation{Department of Astronomy, The Ohio State
    University, Columbus, OH, USA}}

\newcommand{\CCAP}{\affiliation{Center for Cosmology and Astroparticle
    Physics, The Ohio State University, Columbus, OH, USA}}

\newcommand{\LPL}{\affiliation{Lunar and Planetary Laboratory,
    University of Arizona, Tucson, AZ 85721 USA}}

\newcommand{\ERAU}{\affiliation{Department of Physical Sciences,
    College of Arts and Sciences, Embry-Riddle Aeronautical
    University, 1 Aerospace Boulevard, Daytona Beach, FL 32114 USA}}
  
\newcommand{\Granada}{\affiliation{Dept. Fısica Teorica y del Cosmos, University of Granada, 18071 Granada, Spain}}
\newcommand{\gone}{\affiliation{Deceased}}

\newcommand{\LMJU}{\affiliation{
Astrophysics Research Institute, Liverpool John Moores University, 146 Brownlow Hill, Liverpool L3 5RF, UK}}

  \newcommand{\MPI}{\affiliation{Max-Planck-Institut fur Astrophysik, Karl-Schwarzschild Stra{\ss}e 1, 85748 Garching, Germany}}

\newcommand{\nextinstitute}{\affiliation{Put the institute of the new author here}}

\author[0000-0001-5393-1608]{E.~Baron}
\email{ebaron@psi.edu}
\PSI
\HS

\author[0000-0002-7566-6080]{J. M. DerKacy}
\email{jmderkacy@stsci.edu}
\STSci

\author[0000-0002-5221-7557]{C. Ashall}
\email{cashall@hawaii.edu}
\IFA

\author[0000-0002-9301-5302]{M.~Shahbandeh}
\email{mshahbandeh@stsci.edu}
\STSci

\author[0000-0002-8077-5572]{Peter H. Hauschildt}
\email{yeti@hs.uni-hamburg.de}
\HS

\author[0000-0002-7129-3002]{T. Barman}
\email{barman@lpl.arizona.edu}
\LPL

\author[0000-0002-4104-7580]{J. P. Aufdenberg}
\email{jason.aufdenberg@erau.edu}
\ERAU

\author[0000-0003-4625-6629]{C.~R.~Burns}
\email{cburns@carnegiescience.edu}
\Carnegie

\author[0000-0002-4338-6586]{N.~Morrell}
\email{nmorrell@carnegiescience.edu}
\LCO

\author[0000-0002-5571-1833]{M.~D.~Stritzinger}
\email{max@phys.au.dk}
\Aarhus

\author[0000-0002-4338-6586]{P.~Hoeflich}
\email{phoeflich77@gmail.com}
\FSU

\author[0000-0001-7186-105X]{K. Medler}
\email{kyle.medler@sky.com}
\IFA
\VT

\author[0009-0001-9148-8421]{E.~Fereidouni}
\email{ef22g@fsu.edu}
\FSU

\author[0000-0002-7305-8321]{C.~M.~Pfeffer}
\email{cmpfeffer@vt.edu}
\IFA
\VT
\GRFP

\author[0000-0001-5888-2542]{T.~Mera}
\email{tycomera@gmail.com}
\FSU

\author[0000-0003-3953-9532]{W.~B.~Hoogendam}
\IFA
\GRFP
\email{willemh@hawaii.edu}

\author[0000-0001-6107-0887]{S. Shiber}
\FSU
\email{sshiber1@lsu.edu} 

\author[0000-0001-6272-5507]{P. J.~Brown}
\email{grbpeter@yahoo.com}
\TAMU

\author[0000-0001-5965-0997]{Divya Mishra}
\email{dimi_24@tamu.edu}
\TAMU

\author[0000-0002-3827-4731]{Inma Dominguez}
\email{inma@ugr.es}
\Granada
\gone

\author[0000-0002-1296-6887]{L. Galbany}
\email{lluisgalbany@gmail.com}
\ICE
\IEEC

\author[0000-0001-6876-8284]{Paolo Mazzali}
\email{P.Mazzali@ljmu.ac.uk}
\LJMU
\MPI

\author[0000-0003-1039-2928]{E.~Y.~Hsiao}
\email{yichi.hsiao@gmail.com}
\FSU

\author[0009-0006-9436-7197]{Huangfei Xiao}
\email{xiaohf.f2023sec@gmail.com}
\FSU

\author[0000-0001-6069-1139]{T. de Jaeger}
\email{dejaeger.thomas@gmail.com}
\LPNHE

\author[0000-0001-8367-7591]{S. Kumar}
\email{sahanak@gmail.com}
\UVA

\submitjournal{ApJ}

\received{\today}
\revised{\today}
\accepted{\today}

\correspondingauthor{Eddie Baron}
\email{ebaron@psi.edu}

\begin{abstract}
We report \jwst spectra and photometry of the underluminous \sniax
2024vjm obtained \phaseexprf restframe days post-explosion. The
spectrum exhibits a rich set of forbidden lines from low-ionization,
intermediate-mass, and iron-group elements, notably the [\ion{Ni}{2}]
6.64~\microns resonance line, which is a direct indicator of stable
nickel. Strong CO
and SiO emission is detected alongside a warm dust continuum; the
spectral properties are consistent with pre-existing rather than newly
formed dust.  Synthetic spectra were computed with the generalized
stellar atmospheres code \phx using simplified ejecta models. The models
reproduce the overall spectral energy distribution and the molecular
emission features reasonably well, but substantially underestimate the
strength of the mid-infrared atomic forbidden lines, leaving the
synthetic spectrum dominated by molecular emission. Experiments in
which the molecular opacity is suppressed do not recover the forbidden
lines; instead, the emission peak migrates to Co and Fe transitions
near 2~\microns. We attribute this discrepancy to poorly constrained
collisional rates and possibly to an excess of iron-group material in
the current ejecta models.  A prominent feature at 12.8~\microns is
not well accounted for by the [\ion{Ne}{2}] 12.81~\microns line,
indicating that the 12.8~\microns feature may be largely due to [\ion{Fe}{3}]. The
presence of CO, SiO, and stable nickel together with the
non-detection of neon places tight constraints on the total ejecta
mass and the nucleosynthetic yields of \sneiax progenitor systems.
\end{abstract}

\section{Introduction}
\label{sec:introduction}

Type Iax supernovae (\sneiax) are a not-uncommon subclass of thermonuclear
supernova \citep{Foley:2013}.
The overall rate of \sneiax is $\sim 5-15$\% of the Type Ia supernovae (\sneia) rate
\citep{Srivastav:2024,Desai:2026}.
\sneiax are easily distinguishable from the
primary class  by their relatively
faint lightcurves and the low expansion velocities observed in their
spectra \citep[see][for a review]{Jha:2017}.

\sneiax have $i$-band maxima ($i_\text{max}$) that range from about
$-18.5$~mag for SN 2012Z to $i_\text{max} \approx -13.7$~mag for \vjm.  \sneiax are typically divided into three categories based upon
their peak brightness \citep{Jha:2017}. Dim \sneiax have $i_\text{max}
> -15.5$~mag, other examples include SN 2008ha, SN 2010ae, and 2019gsc  
\citep{Foley:2009,Stritzinger:2014,Tomasella:2020}.
Intermediate brightness \sneiax have $i_\text{max}
> -18.0$~mag, examples include  SN~2005hk, SN~2019muj,
SN~2022xlp, \pxl, and SN~2025qe
\citep{Phillips:2007,Sahu:2008,Barna:2021,Banhidi:2025,Singh:2025,Magee:2025}. Bright
\sneiax have $i_\text{max} < -18$~mag, members of the bright class include SN~2002cx, SN~2012Z, SN~2014dt,
and SN~2024bfu \citep{Li:2003,Stritzinger:2015,Singh:2018,Magee:2025}.

A variety of models have been proposed to explain the origins of SNe~Iax.
One model is a close binary system  consisting of a  
C+O+Ne white dwarf (WD) with a helium donor companion star
\citep{Fink:2014,Kromer:2015,Feldman:2023}. Helium accretion induces
thermonuclear burning in the primary, that is, a deflagration, but the 
WD primary
fails to fully disrupt, resulting in a bound remnant. This model is
known as the failed deflagration model.
Another model is the merger of an O+Ne WD with a C+O WD
 producing a failed deflagration of the C+O WD,
ejecting a very small amount of mass, producing an event that is even
dimmer than \vjm \citep{Kashyap:2018}.
Finally, the pulsating delayed
detonation (PDD) of  a near Chandrasekhar-mass ($M_{Ch} \sim 1.4 M_{\odot}$)  WD
\citep{Stritzinger:2015} may also explain SNe~Iax.

SNe~Iax  are quite diverse in peak luminosity, which is expected
to vary with progenitor system and explosion mechanism.
The failed deflagration 
models of C+O+Ne WDs tend to produce enough \Nifs to power the
explosion; however, their ejecta
mass $0.01 < M_\text{ejecta} < 0.1$~\Msun and kinetic energies $10^{48} <
E_\text{KE} < 10^{49}$~erg are too low \citep{Feldman:2023} and hence the models do not fit the post-maximum light curve evolution. The merger
of an O+Ne WD and a C+O WD discussed above
produces extremely faint and fast
light curves, this failing may be alleviated by
more massive O+Ne primaries \citep{Kashyap:2018}.
It has been argued \citep{Neopane:2022}, that WD mergers naturally produce
highly magnetized, uniformly rotating WDs that contract and spin up,
leading to both: near-$M_{Ch}$  deflagration to detonation
transition \citep[DDT][]
{Khokhlov:1991_a,Khokhlov:1991_b,Arnett:1994_a,Arnett:1994_b,Hoeflich_etal_1995,Hoeflich:1995,Wheeler:1995,Hoeflich:1996,Poludnenko:2019}
producing normal \sneia; and pure deflagrations leading to \sneiax.
The merger models of \citet{Neopane:2022} result in central detonations,
unlike what is found in \jwst studies of normal bright \sneia \citep{DerKacy:2023,DerKacy:2024,Ashall:2024,Kwok:2026}.

All the failed deflagration models described above leave behind a
bound remnant.
Late‑time observations of SN 2012Z revealed a flux excess
between 500–1400 days \citep{McCully:2022}. \citet{Schwab:2025}
extended the baseline to 2500 days with HST,  concluding that the
excess is due to a
bound remnant. SN 2014dt showed early dust signatures
\citep{Fox:2016}, but lacked enough data to pinpoint the dust's
origin.
It could well be that dust is produced, but
with different compositions, for example: carbonates in dim; silicates in
intermediate; and iron dust in bright \sneiax. 

\section{\vjm}
\label{sec:vjm}

\vjm was discovered on September 13, 2024 (MJD 60566) by the BlackGem
telescope \citep{Tranin:2024} with a $q$-band  apparent magnitude of
18.8 \citep[$q$ is a wide band
$420-720$~nm filter][]{Groot:2024}. The
supernova is located in the nearby galaxy NGC~6744, at an estimated
distance of 9.4 Mpc \citep{Anand_2021,Zimmerman_2026_vjm}.
\vjm is the  closest  SN Iax ever discovered.\footnote{Despite SN~2022xlp's
smaller redshift, \vjm's distance is 
$9.4 \pm 0.4$~Mpc  while SN~2022xlp's distance is 22.1 Mpc \citep{Banhidi:2025}.}
\vjm is also the dimmest with $\imax \sim -13.7$~mag
\citep{Zimmerman_2026_vjm}.
The Asteroid
Terrestrial-impact Last Alert System 
\citep[ATLAS][]{Tonry:2018} light curve implies a value of
$M_{o,\text{peak}} = -13.6 \pm 0.4$~mag and $t_\text{peak} \sim $ MJD 60575.6.
We take the time of $B_\text{max}$ as
\tbmax = MJD
60572.3 and the risetime as 6.6~days, corresponding to \texp = MJD
60565.6 \citep{Zimmerman_2026_vjm}.
The foreground reddening to NGC~6744 is  $A_V = 0.118$~mag or E(B-V) =
0.038 \citep{Schlafly:2011}. Since \vjm is on the edge of a near face-on galaxy, we neglect
any reddening due to the host.

\vjm  was observed by \jwst  $\sim +$12d after peak with
NIRSpec and MIRI/MRS \citep{Kwok:2025}.  With the use of MIRI/MRS,
their spectrum extends from $ 0.9 \la \lambda \la 20$~\microns. They
identify [\ion{Mg}{2}] 4.76~\microns, [\ion{Mg}{2}] 9.71~\microns, [\ion{Ne}{2}]
12.81~\microns, and \ion{O}{1} 2.76~\microns, which they attribute to
unburned material. They find that the ejecta is well mixed, indicating
a deflagration and that the spectra are well fit with failed
deflagration models, but that they require extra emission in the MIR,
which points to a bound remnant.

We obtained Director's discretionary time to observe \vjm at around
200~days
through program JWST-GO-9231
\citep[PI: Baron;][]{Baron_2025_vjm}.  The purpose of this program is
to obtain  NIR and MIR spectroscopic and photometric data to probe line
formation,  the
formation of molecules,  and  the formation of dust in \vjm. \autoref{tab:vjm_details}
provides the basic observational parameters of \vjm adopted in this work.

\section{Observations and Data Reduction}
  \label{sec:data}

  \begin{table}
    \centering
    \caption{Basic observational parameters of \vjm used in this work. \label{tab:vjm_details}}
    \begin{tabular}{ccc}
    \hline
    \hline
    Parameter & Value & Source \\
    \hline 
    RA & 19$^{h}$09$^{m}$25.780$^{s}$ & (1)  \\
    DEC & $-$63\arcdeg 50  \arcmin 01.7532\arcsec & (1) \\ 
    $T_\text{exp}$ (MJD) & $60565.6 \pm 1.0$ & (2) \\
    $M_{o_\text{max}}$ (mag) & $\sim-13.8$ &  \\
    $z$ & 0.0028 & (1) \\
    Distance (Mpc) & $9.39  \pm 0.43$ & (3,4) \\
    $\mu$ (Mpc) & $ 29.86\pm 0.01 $ & (3,4) \\ 
    $E$(B-V)$_\text{MW}$ (mag) & $0.038\pm 0.02$ & (5) \\
    $E$(B-V)$_\text{Host}$ (mag)& $0$\\
    Host & NGC 6744 & (1)  \\
    \hline
    \end{tabular}
    \tablerefs{(1) \href{TNS}{https://www.wis-tns.org/object/2024vjm},
    (2) \cite{Kwok:2025}, (3) \cite{Anand_2021} (4) \cite{Zimmerman_2026_vjm}, (5) \cite{Schlafly:2011}}
  \end{table}

 \subsection{JWST Spectra}

 Spectroscopic observations were obtained at \phasebmax~days past maximum 
 (\phasebmaxrf~days in the restframe, which we will use throughout)
 using the fixed slit with the Near Infrared Spectrograph (NIRSpec)
 and the  Mid-Infrared Instrument (MIRI) low-resolution spectrometer
 (LRS). This produced continuous spectral data from
 $\sim$0.7-14~$\micron$.
The resolving power of these observations is $\sim$1000 for NIRSpec and
$\sim$100 for MIRI/LRS. 
 All data were reduced using the \jwst pipeline version 1.18.0
 \citep{bushouse_2025_15178003} and
 CRDS context \texttt{jwst\_1371.pmap}. A log of the spectral
 observations can be found in \autoref{tab:JWST_spec_info}.

 \subsection{JWST Imaging}
We obtained simultaneous imaging of \vjm in the \jwst MIRI filters
\textit{F1500W}, \textit{F1800W}, and \textit{F2100W}.  The data were
reduced using the same method described in \citet{Medler:2025_ixf}.
Point-spread function (PSF) photometry was performed using a custom
Python notebook built on the WebbPSF-based \textit{space\_phot} package
\citep{Perrin:2014}. We select the PSF width to minimize
residuals in the fit, accounting for the expected wavelength-dependent
broadening. For each dither in which \vjm is detected, we
average the individual flux measurements and estimate the photometric
uncertainty from the standard deviation across the dithers.  A log of
the photometric observations is presented in \autoref{tab:JWST_phot}.

\begin{deluxetable}{cc}
  \tablecaption{Log of the \jwst spectral observations.
  \label{tab:JWST_spec_info}} 
  \tablehead{\colhead{Parameter} & \colhead{Value}  }
  \startdata
    \hline
    \multicolumn{2}{c}{NIRSpec Spectral Observations} \\
    \hline
    Mode & Fixed Slit  \\
    $T_{\rm obs}$ (MJD) & 60768.86/60768.94\\
    Restframe Phase wrt $B_\text{max}$ (days) & +\phasebmaxrf\\ 
    Restframe Phase wrt explosion (days) & \phaseexprf\\ 
    Slit & S400A1 \\
    Subarray & SUBS400A1 \\
    Grating-Filter & G140M-F070LP/G235M-F170LP/G395M-F290LP \\
    Exp Time (s) & 238.44/331.92/378.66 \\
    Groups per Integration & 50/70/80 \\
    Integrations per Exp. & 1/1/1 \\
    Total Dithers & 3/3/3 \\
        Total Integrations & 3/3/3 \\
    Readout Pattern & NRSRapid \\
    \hline
    \multicolumn{2}{c}{MIRI Spectral Observations}  \\
    \hline
    Mode & LRS \\
    $T_{\rm obs}$ (MJD) & 60768.94/60769.09  \\
    Restframe Phase (days) &  +\phasebmaxrf \\  
    Groups per Integration & 100 \\
    Integrations per Exp. & 20 \\
    Exposures per Dither & 1 \\
    Total Dithers & 2 \\
    Total Exp Time (s) & 11,205.61
  \enddata
\end{deluxetable}

\begin{deluxetable}{ccccc}
\tabletypesize{\footnotesize}
\tablewidth{\columnwidth}
\tablecaption{The photometry obtained with \jwst.\label{tab:JWST_phot}}
\tablehead{
  \colhead{Filter} & \colhead{Obs. date} & \colhead{Exp. time} & \colhead{Phase\tablenotemark{a}} &
  \colhead{Magnitude} \\
  \colhead{} & \colhead{(MJD)} & \colhead{(s)} & \colhead{(days)} &
  \colhead{(Mag)}
}
\startdata
\textit{F1500W} & 60769.09/60769.12 & 1110.0 & +196.83 & $19.64 \pm 0.01\phn$ \\
\textit{F1800W} & 60769.12/60769.20 & 1744.4 & +196.91 & $19.78 \pm 0.02\phn$ \\
\textit{F2100W} & 60769.12/60769.20 & 4191.4 & +196.91 & $20.18 \pm 0.02\phn$ \\
\enddata
\tablenotetext{a}{Restframe days with respect to B$_\text{max}$
  MJD 60572.29.}
\end{deluxetable}

\subsection{SED}
\begin{figure*}[ht]
    \centering
    \includegraphics[width=\textwidth]{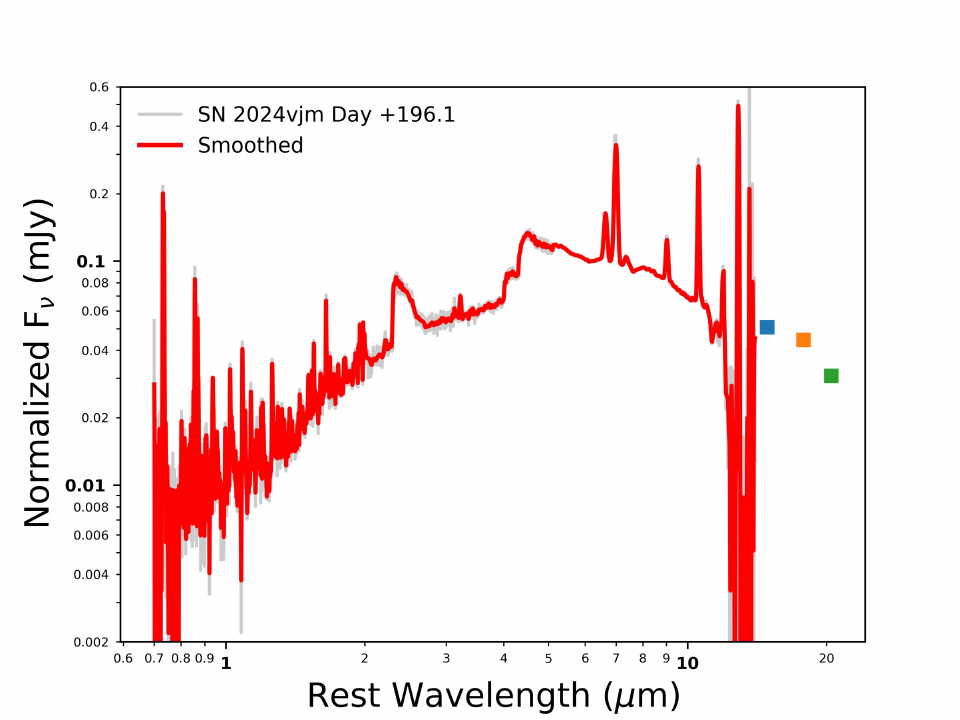}
    \caption{SED of \vjm. The raw and smoothed Day +\phasebmaxrf \vjm
      \jwst spectrum 
plus the day 196.9 JWST photometry (shown by squares).}
    \label{fig:sed}
\end{figure*}

\autoref{fig:sed} shows the resulting spectral energy distribution
(SED) including both the spectroscopy and photometry. 
The underlying continuum appears to roughly conform to a
blackbody-like shape, with significant emission from a few forbidden
lines, spread throughout the wavelength region including the
[\ion{Ca}{2}] 0.73~\microns, the [\ion{Ni}{2}] 6.64~\microns line, the
[\ion{Ar}{2}] 6.9~\microns line, the [\ion{Co}{2}] 10.52~\microns
line, and the [\ion{Fe}{3}] 12.84~\microns line. There is also clear
emission from both the  CO fundamental and first overtone bands as
well as from the SiO first overtone. This is to our knowledge the
first clear detection of CO and SiO in \sneiax or \sneia.  In the case
of the peculiar 2003fg-like SN LSQ14fmg, \citet{Hsiao:2020_lsq14fmg} 
speculated that a steep drop in the  light curve  about a month post
maximum was due to the 
  formation of CO and the cooling it provides, but there was no
  spectroscopic confirmation of the presence of CO. CO has
  been observed in numerous core collapse events including SNe 1987A,
  2023ixf, and 2024ggi \citep{Wooden_1993,Medler:2025_ixf,Mera_2026} and
  SiO has been observed in SNe 2004et and 2024ggi \citep{Kotak_2009,Mera_2026}. Dust has been
  observed to form in SN~2018cvt \citep{Wang_2024_18cvt}, but
  SN~2018cvt is a SN~Ia CSM,  a rare subclass of \sneia, and also
  in the 2003fg-like SN~2022pul 
  \citep[][J.~Johansson, in preparation]{Siebert:2024,Kwok:2024_pul}. Spitzer detected a MIR excess in the \sniax
  2014dt about one year post-explosion \citep{Fox:2016}. This has
  been interpreted as due to newly-formed dust, pre-existing dust, or
  a bound remnant \citep{Szalai:2019,Fox:2016,Foley:2016}. In neither SN~2018cvt nor SN~2022pul, was
  molecule formation of either CO or SiO observed. In \vjm, the
long wavelength ($15 < \lambda < 21$~\microns)  photometry does not show any evidence for significant
emission from cold ($T \la 500$~K) dust. 

\begin{figure*}[ht]
    \centering
    \includegraphics[width=\textwidth]{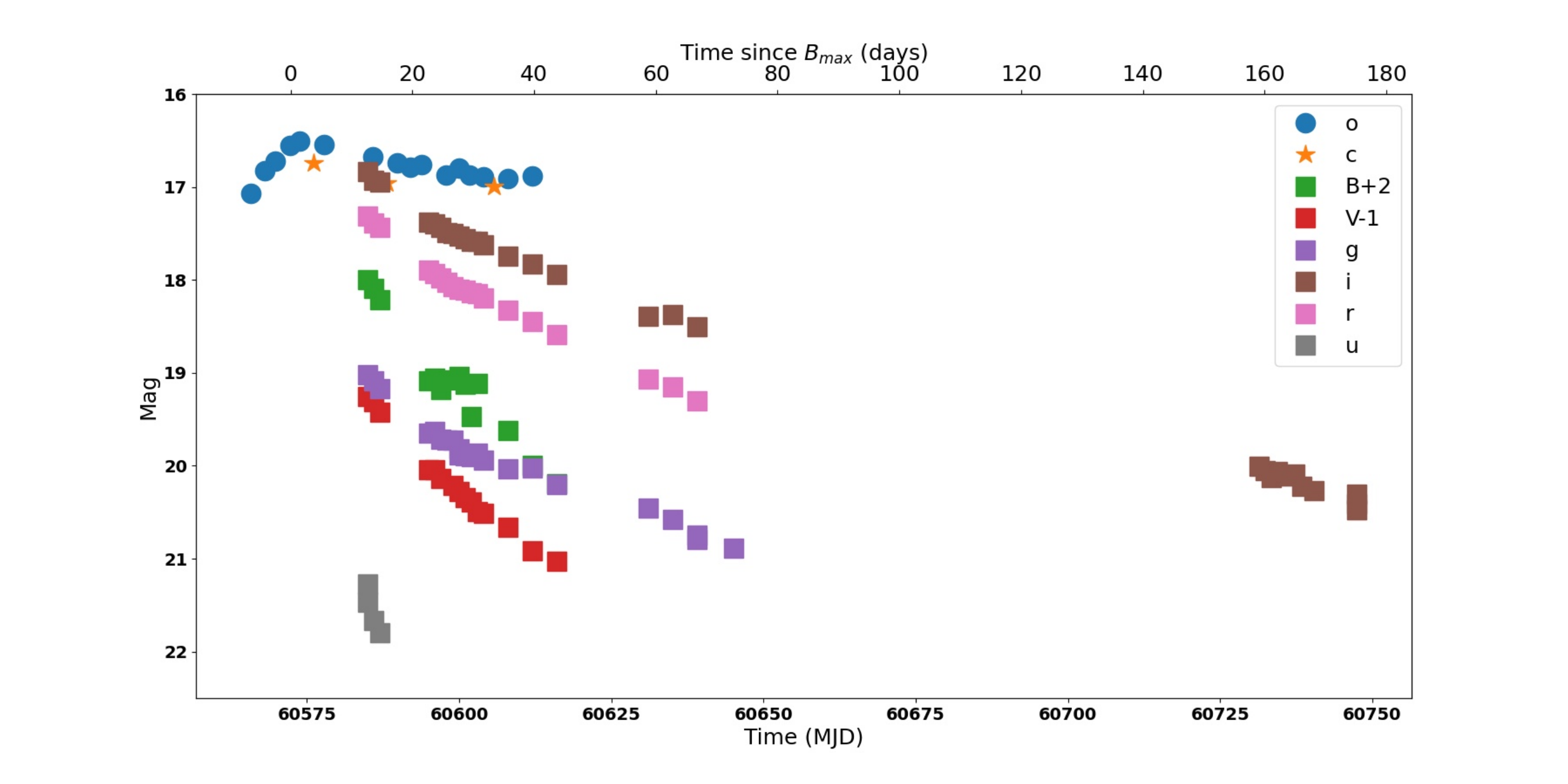}
    \caption{The \vjm light curve obtained by ATLAS and POISE. 
    }
    \label{fig:lc}
\end{figure*}

\citet{Zimmerman_2026_vjm} present extensive light curve coverage of
\vjm, but we present additional photometry.  \autoref{fig:lc} shows the light curve of \vjm obtained by ATLAS  and by the Precision Observations of Infant Supernova Explosions  \citep[POISE][]{Burns:2021}. 
From the ATLAS $o$-band light curve we find that $t_{\text{o}_\text{max}}=$~MJD 60575.6,
$o_\text{max} = 16.5$~mag,  and  $o_\text{max} = -12.9$~mag
correcting for foreground extinction.  The time of maximum light in the $o$-band and our
adopted explosion time lead to a rise time in the $o$-band of 11~days,
with the $o$-band light curve peaking $\sim 4$~days after the $B$-band light curve.

\begin{figure*}[ht]
    \centering
    \includegraphics[width=\textwidth]{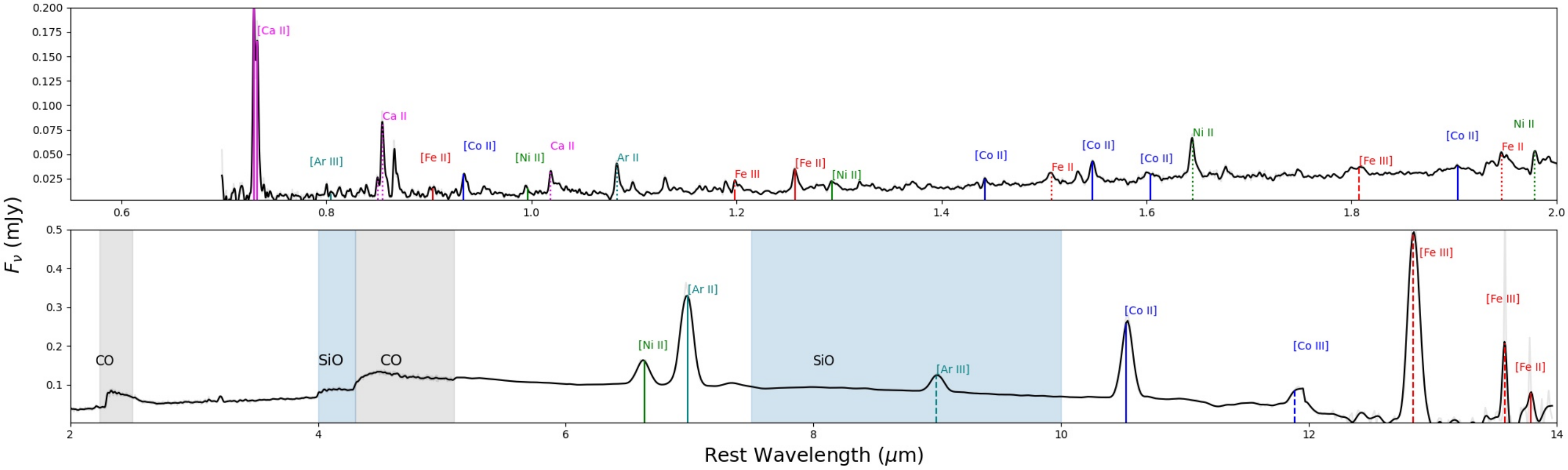}
    \caption{The observed Day +\phasebmaxrf \vjm \jwst spectrum, with tentative
      line identifications. The CO and SiO fundamental and first
      overtone bands are shaded \citep{Heras:1999}.}
    \label{fig:quickid}
\end{figure*}

Using the (to be made public) tool QuickID (E.~Baron, in preparation), 
\autoref{fig:quickid} shows line 
identifications. QuickID uses the database of \citet{vanHoof:2018} and
simple line strength criteria, chosen by the user to narrow down the possible line IDs. No
modeling is done and thus, the line IDs are tentative. We
identify forbidden  and permitted lines. We find lines due to intermediate mass elements
[\ion{Ca}{2}], 
[\ion{Ar}{2}], [\ion{Ar}{3}], and iron group elements [\ion{Fe}{2}],
[\ion{Fe}{3}], [\ion{Co}{2}], [\ion{Co}{3}], and [\ion{Ni}{2}]. It is
possible that there is a feature due to [\ion{Ne}{2}], but as
discussed below we don't find strong evidence to support that
identification. 

\begin{figure*}[ht]
    \centering
    \includegraphics[width=\textwidth]{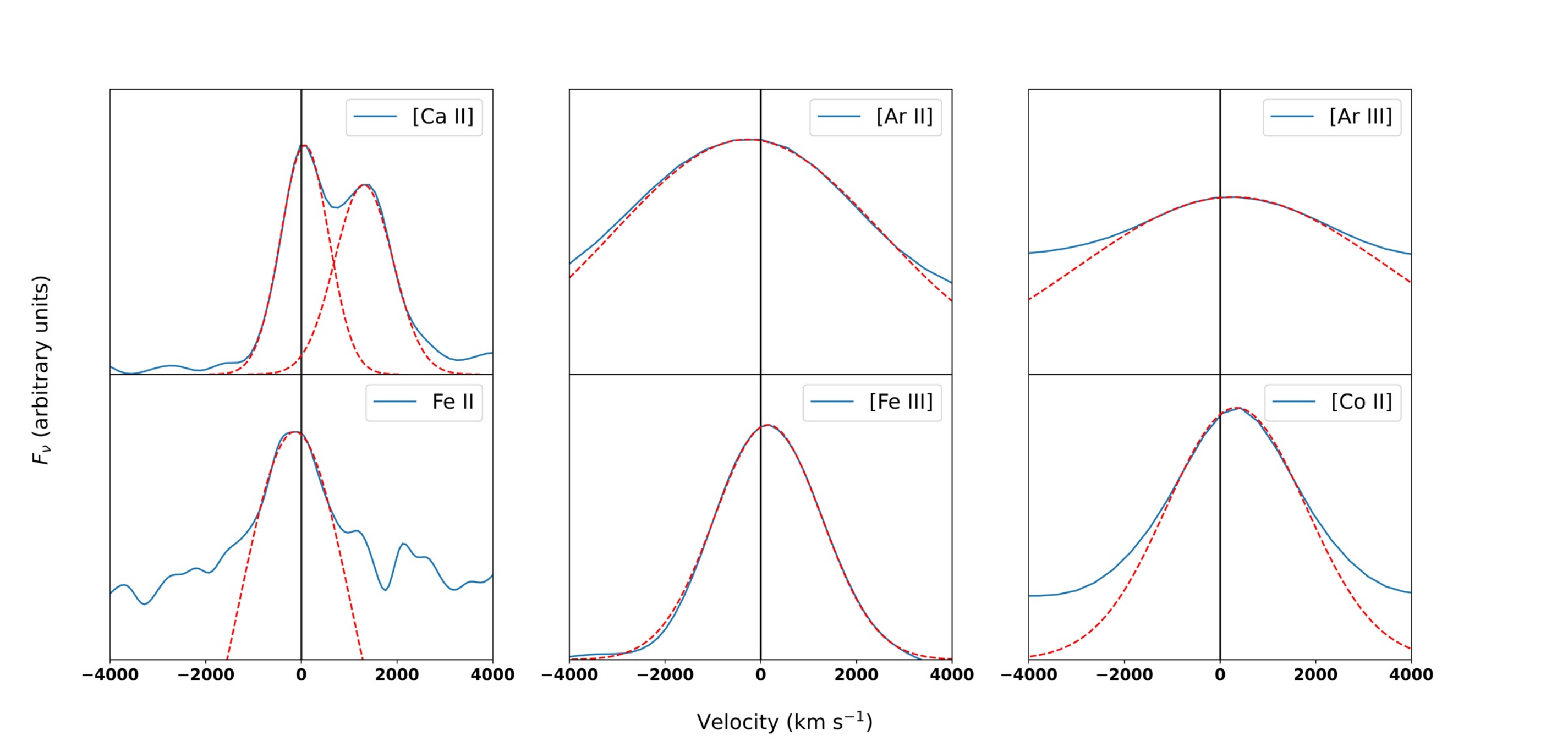}
    \caption{The strong intermediate mass element lines (top row) and
      iron group element lines (bottom row). The blue lines are the
      data and the red dashed lines a Gaussian fit to the feature. The argon
      lines are broader and somewhat more asymmetric in
      velocity space with red or blue shifts. The [\ion{Ca}{2}] doublet
      is resolved, as is the Ca IR triplet. [\ion{Ca}{2}] shows very
      symmetric line shape, while the [\ion{Fe}{3}] has a small redshift
      and the [\ion{Co}{2}] line has a more marked redshift. The
      Gaussian fit results are shown in \autoref{tab:vels}.}
    \label{fig:vel_plot}
\end{figure*}

\autoref{fig:vel_plot} shows the strong lines of both the intermediate
mass elements and the iron group elements in velocity space. The [\ion{Ca}{2}] doublet is resolved as are the strong components of the Ca IR
triplet, indicating low ejecta velocities. The centroid and velocity
width obtained from Gaussian fits to the features are shown in
\autoref{tab:vels}. The argon lines are broader with widths around
3000~\kmps, while the iron group elements are narrower with widths
around 1500~\kmps. However, the argon lines are not flat topped so the
argon is not confined to a shell. This analysis shows that the ejecta
of \vjm is ``almost fully mixed'' or ``almost radially
stratified''. 
 As noted
above, the line
IDs of \autoref{fig:quickid} do not take line blending into account,
thus, in the absence of detailed models it is not possible to draw
strong conclusions about the velocity extent of the broader
lines. However, both the identifications of [\ion{Ca}{2}] and the \ion{Ca}{2} IR
triplet are solid. These calcium features show rather symmetric
profiles in velocity space, while the
 [\ion{Co}{2}] feature shows a slight redshift.
Since the [\ion{Co}{2}] line is a
resonance line, it should be relatively immune to line blending, the
shift could due to the density and ionization profile
or it could be due to
a true asymmetry in the elements produced in later stage burning and
the calcium could be primordial. 
The narrow width and near zero redshift of both components of the
[\ion{Ca}{2}] doublet are interesting. It seems unlikely that the
calcium emission is from below that of the iron group elements,
suggesting that  calcium originates in a wind
external to the ejecta. A $\sim 500$~\kmps wind
is not a feature of models for the formation of
\sneiax. \citet{Siebert:2023} found extremely narrow [\ion{Ca}{2}]
with a width of around 200~\kmps, in the 2003fg-like SN~2020hvf (as well
as other 2003fg-like SNe), which they attributed as due to the wind from
the intact companion star. They also found a broad [\ion{Ca}{2}]
feature from the ejecta.

\begin{deluxetable}{lrrr}
\tabletypesize{\footnotesize}
\tablewidth{\columnwidth}
\tablecaption{
  Widths and Central Velocities of Lines shown in
  \autoref{fig:vel_plot}.
  \label{tab:vels}}
\tablehead{
  \colhead{Feature} &\colhead{Rest Wavelength} & \colhead{Center} & \colhead{Width}\\
  \colhead{} &\colhead{(vacuum)} & \colhead{(\kmps)} & \colhead{(\kmps)} 
}
\startdata
    {[Ca II]}& 0.7291~\microns& 70 & 500 \\
    {[Ca II]}& 0.7293~\microns& 30 & 540\\
    {[Ar II]}& 6.985~\microns& $-$270& 2800\\
    {[Ar III]}& 8.991~\microns& 270 &  3200\\
    Fe II &1.507~\microns& $-$140& 1200\\
    {[Fe III]} &12.841~\microns& 150&  1100\\
    {[Co II]} &10.522~\microns&340&1500\\
\enddata
\end{deluxetable}

\section{Spectral Comparison}

\begin{figure*}[ht]
    \centering
    \includegraphics[width=\textwidth]{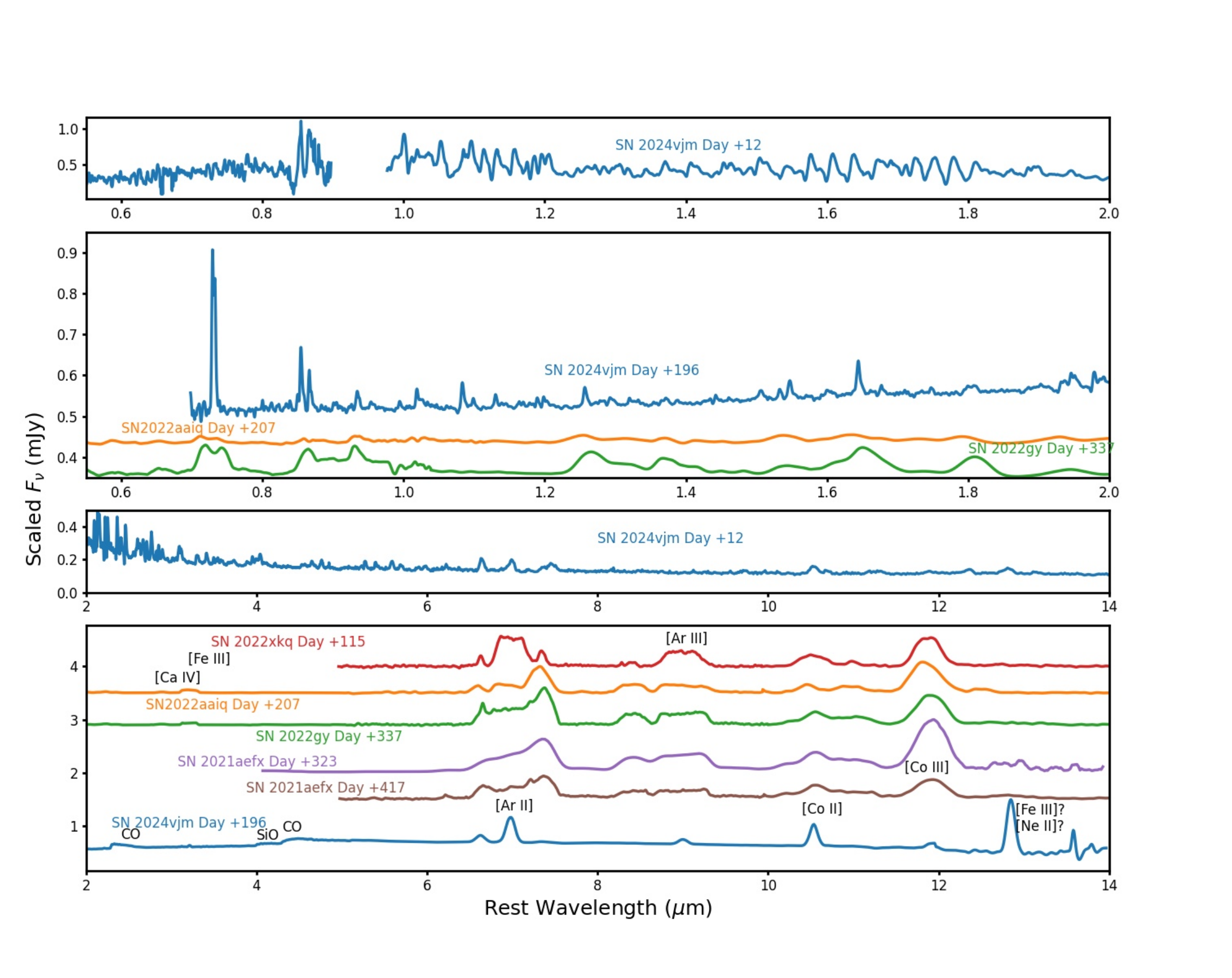}
    \caption{The observed Day +\phasebmaxrf \vjm \jwst spectrum plotted against
      other SNe Ia at similar epochs that have also been observed by
      \jwst. The data for SN 2022xkq are from \citet{DerKacy:2024} as
      rereduced by \citet{Kwok:2025b}, the data for SNe 2022aaiq and
      2022gy are from \citet{Kwok:2025b}, and the data SN 2021aefx are
      from \citet{Ashall:2024}. The thin panels show \vjm at Day +12
      \citep{Kwok:2025}.
      }
    \label{fig:comp_sneia}
\end{figure*}

\autoref{fig:comp_sneia} shows a suite of \sneia (including \sneiax)
that have been observed by \jwst. All phases are with respect to
$B$-band maximum. The 1991bg-like SN 2022xkq spectrum
obtained at day +130 \citep{DerKacy:2024},  is
earlier than that of \vjm, while SNe 2021aefx \citep[days +323, +417][]{DerKacy:2023,Ashall:2024},
2022gy (day +337), and 2022aaiq (day +207) \citep{Kwok:2025b} are normal SNe Ia.  While the ions that make up the spectra of both
the \sneia and \sneiax are similar, the difference in the nature of
the spectra between the \sneia and \vjm is striking. Not only are the
features in \sneia much broader than those present in \vjm, but their
relative strengths are quite different. Whereas the
dominant feature in the NIR for SNe 2022gy and 2022aaiq is due to a
blend of [\ion{Ca}{4}] and [\ion{Fe}{3}] \citep{Kwok:2025b}, in \vjm
it is due to CO.  Both the normal SNe 2022gy and 2022aaiq and
\vjm show
a flat feature in the wavelength range $\approx 4-4.2$~\microns,
attributed to [\ion{Ca}{4}] by \citet{Kwok:2025b}. However, given the
low temperature required for the formation of CO, it seems unlikely
that  this feature is due to such a highly
ionized species, at least in the case of \vjm (see \autoref{sec:disc}
for a discussion of this particular feature).
Moving further to the red, [\ion{Ar}{2}] is prominent in both \vjm and
SN 2022xkq, whereas [\ion{Ar}{3}] forms a prominent flat-topped
feature in the normal \sneia, but is a relatively narrow symmetric
feature in \vjm. The same pattern continues to the red,  where
in the \sneia the strongest feature is the [\ion{Co}{3}] resonance
line at 11.88~\microns, but in \vjm it is the [\ion{Co}{2}]
10.52~\microns resonance line that dominates over the higher
ionization [\ion{Co}{3}] line. Interestingly, further to the red, the
\sneia lack the feature at 12.8~\microns seen in \vjm, which is likely due to the
       [\ion{Fe}{3}] 12.84~\microns line, but could be due to the
       [\ion{Ne}{2}] 12.81~\microns resonance line.

\section{Dust Emission}

\begin{figure*}[ht]
    \gridline{\fig{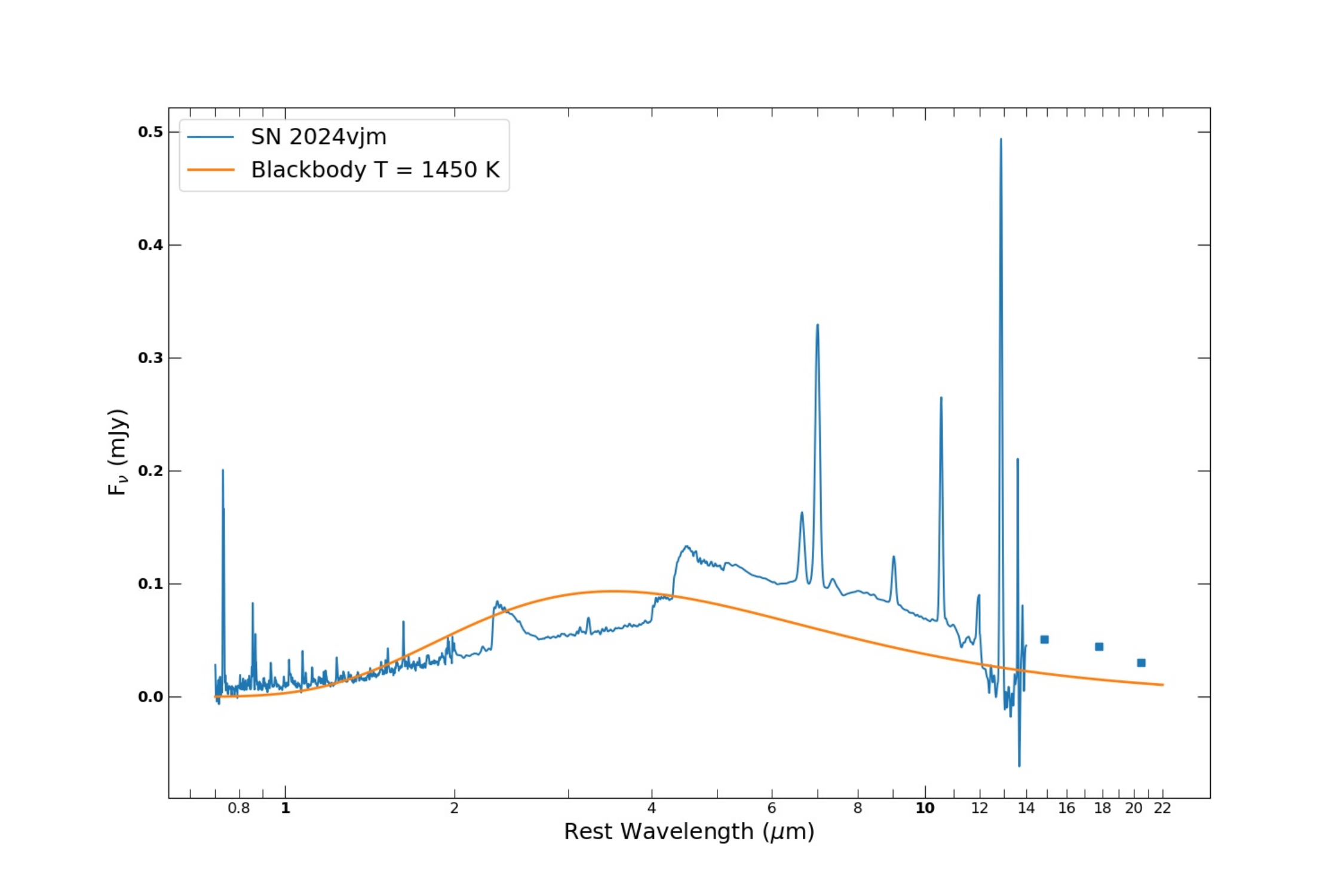}{0.49\textwidth}{(a)}
    \fig{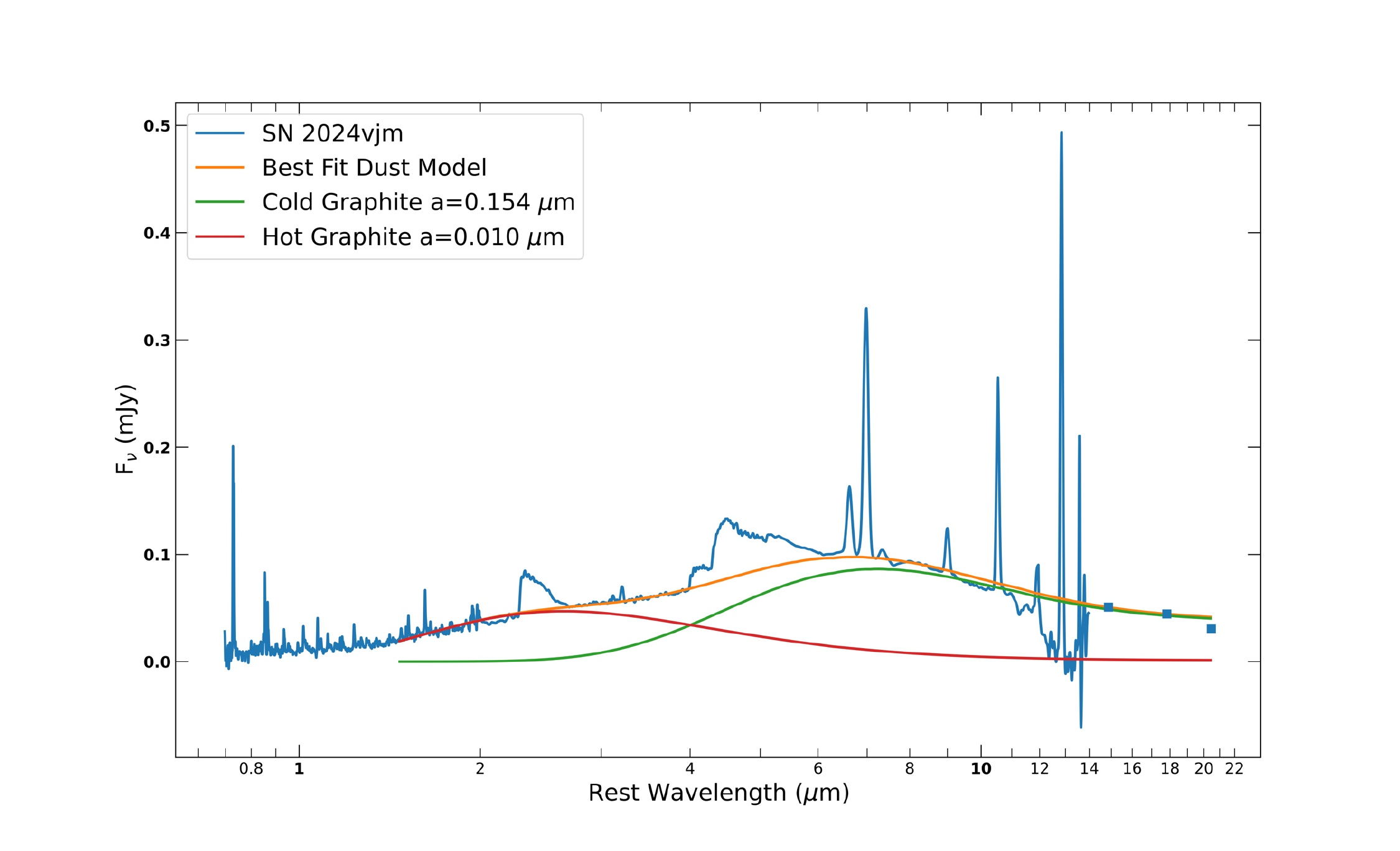}{0.49\textwidth}{(b)}}
    \gridline{\fig{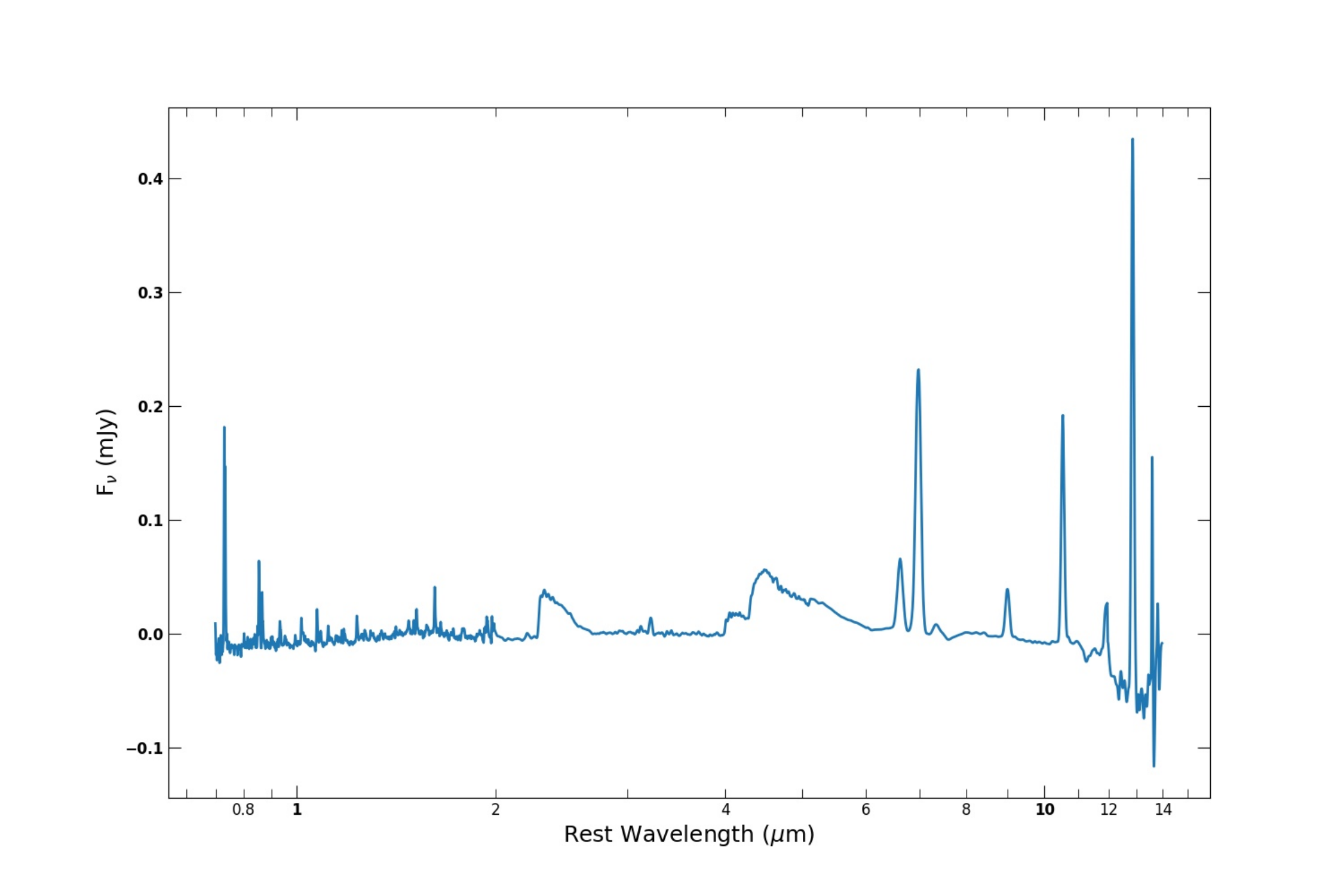}{\textwidth}{(c)}}
    \caption{(a) The observed Day +\phasebmaxrf \vjm \jwst spectrum
      plotted against 
      the the best fit blackbody. (b) The observed Day +\phasebmaxrf \vjm \jwst spectrum plotted against
      the the best fit dust model. The two components of the model
      consisting of cold, 0.15~\microns and hot,
      0.01~\microns graphite are also shown.  (c) The observed Day +\phasebmaxrf \vjm \jwst spectrum with the best
      fit dust model (b) removed.    \label{fig:vjm_bb_dust}} 
\end{figure*}

There is a strong indication of an
underlying dust continuum present in \vjm.
 \autoref{fig:vjm_bb_dust}(a) shows our best fit blackbody to the
\vjm SED, giving a temperature of $T_\text{BB} \approx 1400$~K. With
only a single epoch, it is not possible to ascertain whether this dust
is newly formed or preexisting. However, a simple estimate of the dust
radius gives a $R_d = 2.1 \times 10^{16}$~cm
\citep{Shahbandeh:2025_05ip}, whereas the ejecta, even moving at a
velocity of 10,000~\kmps would have only reached a radius of about
$1.7 (10000/v_\text{max} (\kmps)) \times 10^{16}$~cm, where
$v_\text{max}$ is the true maximum ejecta velocity. As noted in
\citet{Shahbandeh:2025_05ip}, this radius estimate is for the
optically thin case and provides a minimum radius for the dust.
That means the dust may reside in a shell
with a larger radius, but 
 if the radius of the shell were any smaller, the
dust would become optically thick.
This suggests, but does not prove that
the dust is pre-existing rather than newly formed.

To go a little deeper, we calculated a simple optically thin dust
model. The observations were masked to remove the molecular emission. This gives a fit with two graphite components $T \sim 500$~K
and $T \sim 1250$~K and a total dust mass of $M_\text{dust} = 2.41
\times 10^{-4}$~\Msun. \autoref{fig:vjm_bb_dust}(b) shows the best fit
model compared to the \vjm SED, including the photometry points. Clearly, the more physical fit shown
in \autoref{fig:vjm_bb_dust}(b)
gives significantly smaller residuals to the dust continuum than the
simple blackbody assumption of  \autoref{fig:vjm_bb_dust}(a).   \autoref{fig:vjm_bb_dust}(c) shows
the SED with this dust continuum removed. It is this SED that we will
compare to models.

\section{Models}

In order to go beyond the line identifications, we consider some
simple models of \sneiax. While the favored model for \sneiax
  is the failed deflagration model, leaving a bound remnant, radially
  stratified models such as the pulsating delayed detonation
  \citep{Stritzinger:2014} have also been considered. Therefore we
  consider both radially stratified and fully mixed models. For the
radially stratified model we consider a modified version of the
W7 model \citep{Nomoto:1984,Nomoto:1986}  and for the fully mixed
model N1def
\citep{Fink:2014}. Both W7 and N1def are the result of pure
deflagrations, but W7 has radially stratified abundances, which makes
it suitable for our purposes. W7 has a long history of being used as a template for
\sneia
\citep[for example,
see][]{Magee:2024,McCutcheon:2022,Shingles:2020,Cain:2018,Ashall:2016,Mazzali:2014,Thielemann:2007,Jeffery:1992}
and N1def is specifically constructed as a model of faint
\sneiax \citep{Fink:2014}.
We modified W7, taking its abundance and density distributions without
modification, but reducing the total mass and velocity extent. In
order to distinguish this model from W7 itself we refer to it
henceforth as  \wsiax. A similar approach of modifying the kinetic
energy, nickel mass, and total  mass of W7 was taken for modeling
the light curve of the \sniax 2005hk \citep{Sahu:2008}.

To compare the models to data we use the \phxO generalized stellar
atmosphere code \citep{Hauschildt:1999}, the ACES equation of
state and the STOUT model atoms. The STOUT model atoms were
constructed using the data available from \citet{vanHoof:2018}. ACES
\citep{Barman:2011} solves for the abundances assuming  chemical equilibrium in a stellar atmosphere. The method is based upon that of
\citet{Smith_Missen_1982} incorporating new experimental and theoretical thermodynamic data for 839 species
(84 elements, 289 ions, 249 molecules, 217 condensates) with
all available data updates applied. Chemical equilibrium concentrations for all atomic and molecular species (the species are
selected at run time) are determined by the pressure, temperature, and density. This calculation is repeated at each layer in the
atmosphere for each model iteration so that  the final structure is
self-consistent. \phxO has been used on all types of supernovae
\citep{Baron:2025,Baron:2015,Baron:2012,Baron:2006,Baron:2003a,Baron:2003b,Baron:2000,Baron:1996}. The
treatment of gamma-ray deposition and the resulting non-thermal
electrons is described in \citet{Baron:1994}. \phxO was used to
analyze nebular spectra of SN~2011fe and transitional spectra of SNe
2001fe, 2002bo, 2003du, and 2014J \citep{Friesen:2014,Friesen:2017}. The
version and data-sources of \phxO are the same as that presented in \citet{Hauschildt:2025}.

In order to keep the compute time for the models within our means, we
have not solved for the temperature structure, but rather chosen a few
representative structures. Time-dependent effects may be important and
since our treatment of molecules is inherently time-independent, this
is another reason not to expend more computational resources than we
have available to us. It is beyond the scope of the present work to
fully model the dust continuum, therefore we compare the models to the
\vjm SED with the dust continuum subtracted out, that is, with the SED
shown in \autoref{fig:vjm_bb_dust}(c).

\begin{figure*}[ht]
    \centering
    \gridline{\fig{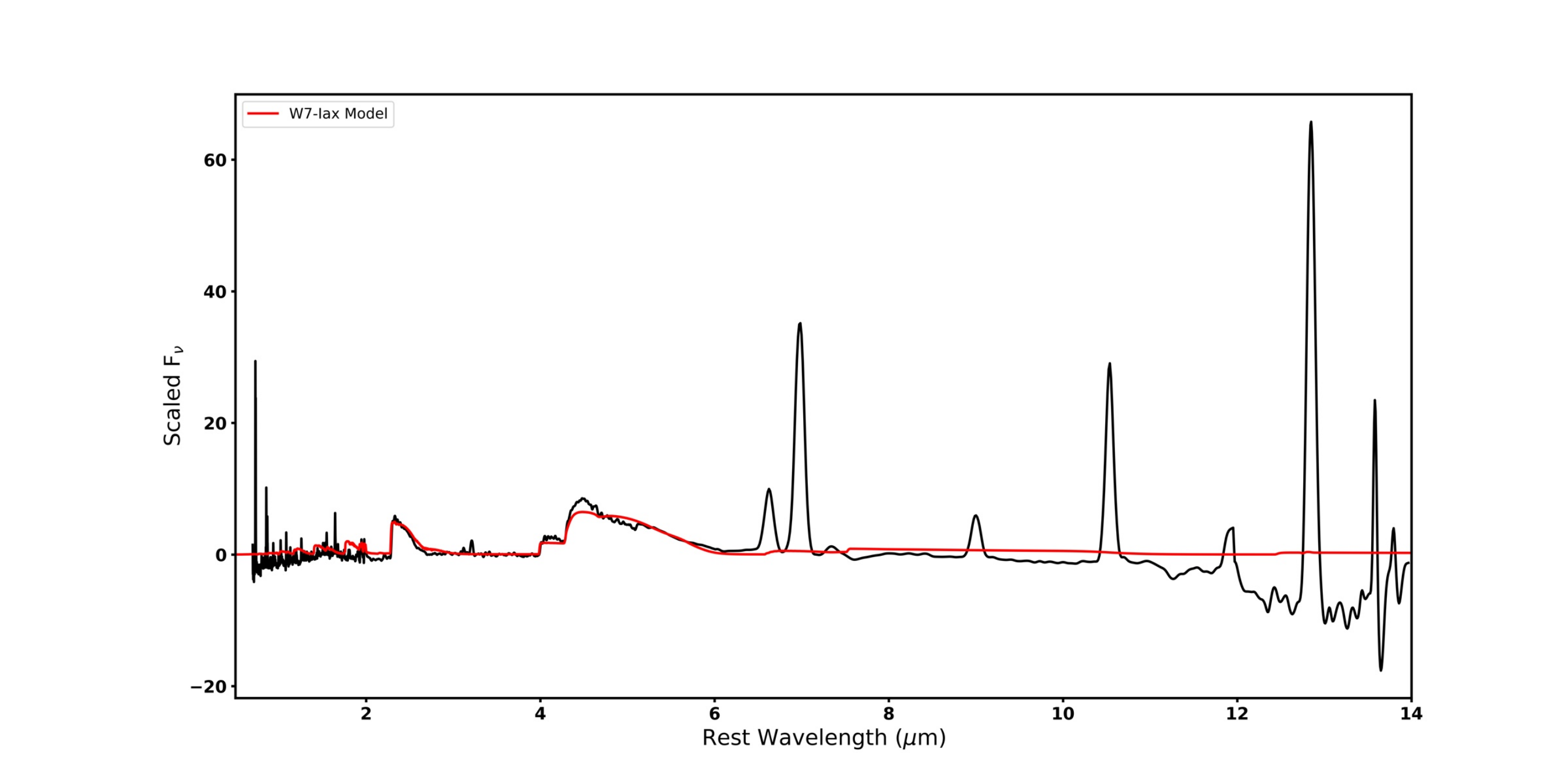}{0.9\textwidth}{(a)}}
    \gridline{\fig{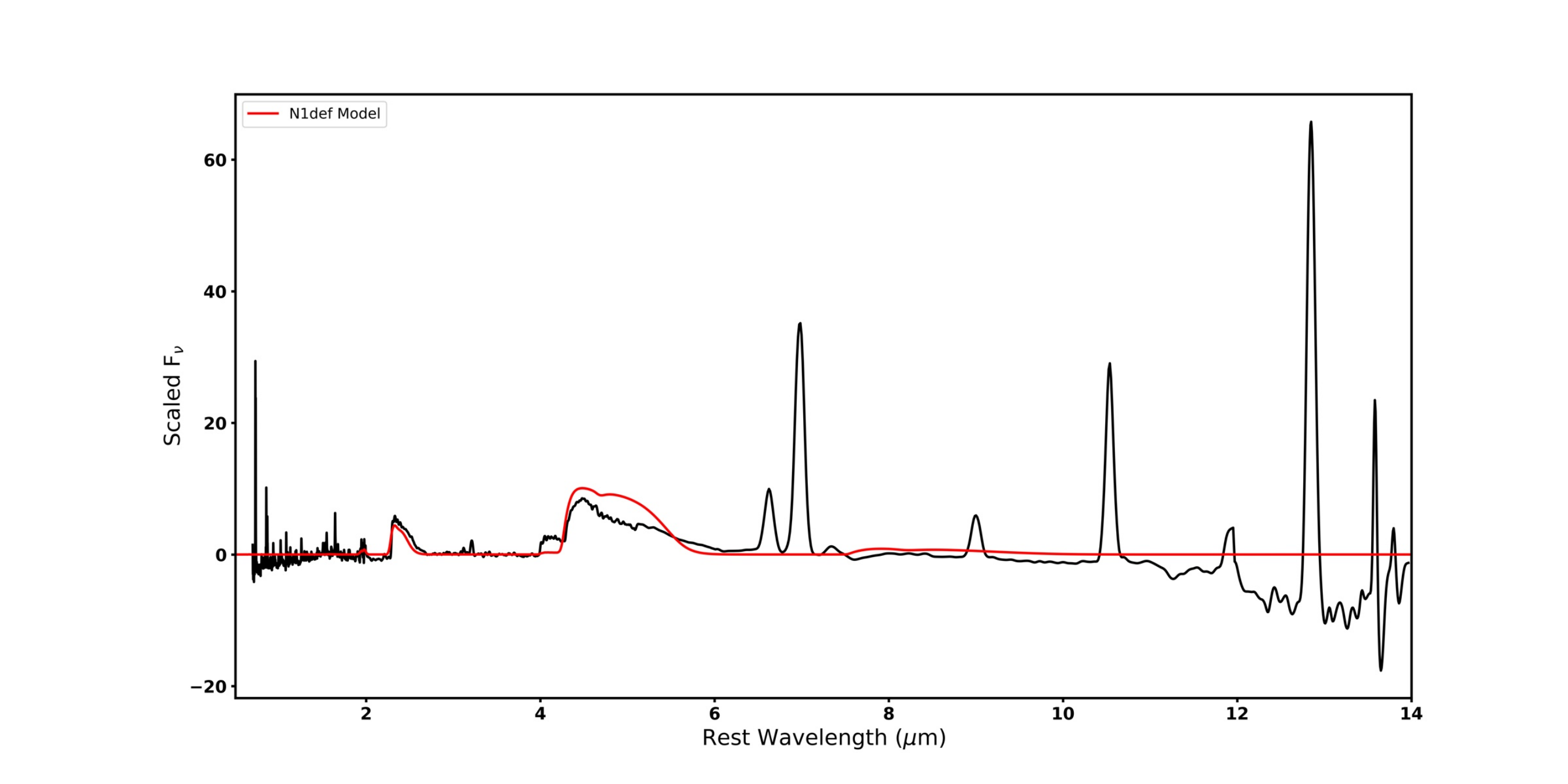}{0.9\textwidth}{(b)}}
\caption{(a) A simple model, \wsiax, using W7 abundances with an ejecta mass of
      0.42~\msol and the velocities reduced by a factor of 10 (the
      \wsiax model). This
      model has an isothermal ejecta with $T=2000$~K, $\gamma$-ray
      deposition, from the (reduced) radioactive nickel is included.
      (b) The N1def model.
       This
      model has an isothermal ejecta with $T=2000$~K and includes
      $\gamma$-ray deposition.
    \label{fig:CO_iso_T2000_wdep}}
\end{figure*}

\autoref{fig:CO_iso_T2000_wdep}(a)  shows the synthetic spectrum of
the \wsiax model.
The \wsiax model is created from W7 by reducing the
density of W7, such that the density profile is unchanged. The
maximum velocity of W7 is reduced by a factor of ten,
 enforcing homology. The resulting \wsiax model has a mass of
 0.42~\Msun,  a maximum velocity of 3000~\kmps and a kinetic energy of
 $3.5 \times 10^{-3}$~foes. The composition
 structure of W7 is unchanged in \wsiax. The model spectrum in \autoref{fig:CO_iso_T2000_wdep}(a) has
an isothermal $T = 2000$~K profile. The model  reproduces the
molecular features of CO with both the 
fundamental band ($\sim 4.3-5.1$~\microns) and the first
overtone ($\sim 2.2-2.5$~\microns) matching the observed spectrum.  There is no striking
evidence for the SiO fundamental ($\sim 7.5-10$~\microns), but the SiO
first overtone ($\sim 4-4.3$~\microns) is clearly evident,  (see \autoref{sec:disc}). The
forbidden lines in the model spectrum are much weaker than those in
the observed SED, with most of the energy emitted by cobalt and iron
transitions near 2~\microns. This could be due to how collisional
rates are determined for the STOUT dataset, but may also be due to many
other factors that are beyond the scope of this work.
We have explored isothermal temperatures in the range
  $1000--3000$~K as well as temperature profiles as obstained by
  \citet{Mera_2026} without significant alteration in the results and
  therefore we take $T=2000$~K as our fiducial model.

\autoref{fig:CO_iso_T2000_wdep}(b) shows the synthetic spectrum of
the N1def model with an isothermal $T = 2000$~K profile.
Like the \wsiax model, N1def reproduces the first overtone of CO,
but the fundamental is not as well reproduced as it is in the \wsiax
model. Again, the forbidden lines are much too weak in the
synthetic spectrum.

\autoref{tab:molecule_mass} shows the basic parameters of the two
models including the mass of CO and SiO produced.

\begin{deluxetable}{lllllll}
  \tablecaption{Model Parameters.  \label{tab:molecule_mass} }
  \tablehead{\colhead{Model} & \colhead{Ejecta Mass} & \colhead{$v_\text{max}$} &
    \colhead{$E_\text{KE}$} & \colhead{E/M} & \colhead{CO Mass} &
    \colhead{SiO Mass}\\
\colhead{} & \colhead{(\Msun)} & \colhead{(\kmps)} &
    \colhead{(foe)} & \colhead{(foe/\Msun)} & \colhead{(\Msun)} &
    \colhead{(\Msun)}  }
  \startdata
  \wsiax & 0.42 & 3000 & $3.5 \times 10^{-3}$ & 0.008 & 0.025 & 0.016\\
  N1def & 0.075 & 8800 & $1.4 \times 10^{-2}$ & 0.018 & 0.023 & 0.0015 \\
  \enddata
\end{deluxetable}

\begin{figure*}[ht]
    \centering
    \gridline{\fig{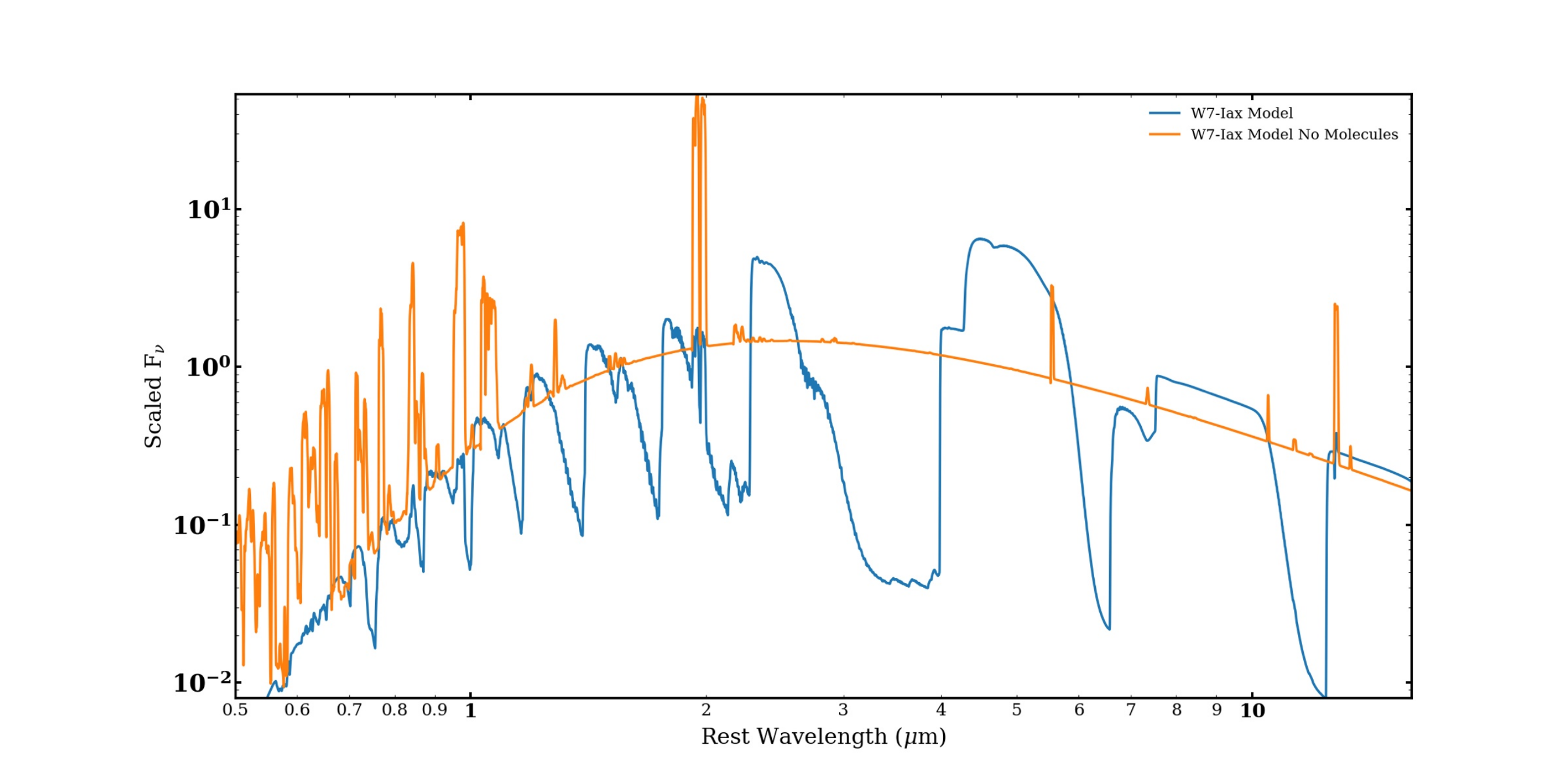}{0.9\textwidth}{(a)}}
    \gridline{\fig{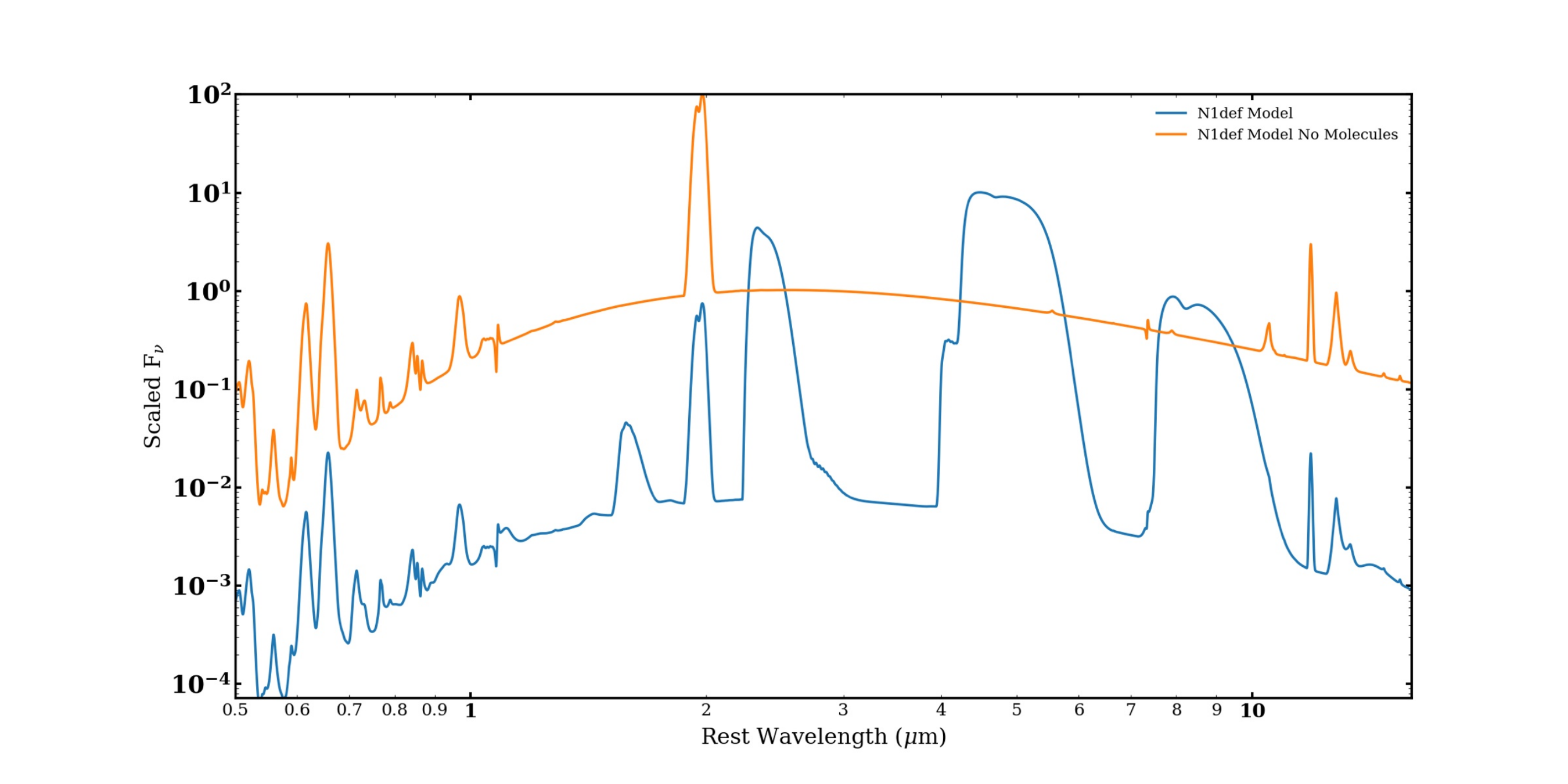}{0.9\textwidth}{(b)}}
    \caption{(a) The \wsiax model shown in
      \autoref{fig:CO_iso_T2000_wdep}(a),
      is compared to the same model but where the  molecule opacity
      has has been ignored. Most of the energy
      comes out in lines due to iron and cobalt near 2~\microns.
      (b) The N1def model shown in
      \autoref{fig:CO_iso_T2000_wdep}(b),
      is compared to the same model but where the  molecule opacity
      has has been ignored. Most of the energy
      comes out in lines due to iron and cobalt near 2~\microns.
    \label{fig:CO_iso_T2000_wdep_nomol}    }

\end{figure*}

 Both the \wsiax model and the N1def model adequately
 reproduce the
 observed \vjm spectrum, blueward of about 6~\microns. In order to examine how well they do
 reproducing the features, the spectra for models were recalculated,
 but the opacity due to molecules was turned off. That is, the
 molecules were still present in the atmosphere, but their opacity was
 neglected.  This is our standard
 method for obtaining unambiguous identifications in the synthetic
 spectra
 \citep{Bongard:2008,DerKacy:2020}. \autoref{fig:CO_iso_T2000_wdep_nomol}
 panels (a) and (b)  
 show the resulting spectra for the \wsiax model and the N1def
 model, respectively. The reason for the lack of strong forbidden
 lines appears to be that most of the energy is emitted by iron and
 cobalt lines
in  the wavelength range around 2~\microns.
\autoref{fig:CO_iso_T2000_wdep_nomol}(a) shows that while the \wsiax
model  has many  optical features, it is
 relatively devoid of atomic features in the NIR. The model
 produces a strong
 feature at 5.54~\microns possibly due to \ion{C}{4}; however, there is no
 corresponding feature in the observed spectrum. The model does
 have a feature at 7.34~\microns likely due to \ion{C}{2} as well as
 the [\ion{Co}{2}] 10.52~\microns line. Interestingly, the model shows
 a feature around 12.8~\microns that appears to be made up of a blend
 of the [\ion{Ne}{2}] 12.81 line and the [\ion{Fe}{3}] 12.84 line,
 with the [\ion{Ne}{2}] portion of the feature not  matching the
 observation, lending credence to the identification of that
 feature as due to [\ion{Fe}{3}] and not [\ion{Ne}{2}].

 \autoref{fig:CO_iso_T2000_wdep_nomol}(b) shows that the N1def
  model also  has most of the observed optical features, although the
  [\ion{Ca}{2}]~0.73~\microns line is very weak (it appears as a
  P-Cygni line in the \wsiax model). Again
 relatively devoid of features in the NIR, this model shows a weak 
 feature at 5.54~\microns possibly due to \ion{C}{4}, that is not in
 the observed spectrum. The model shows
 a P-Cygni feature at 7.34~\microns likely due to \ion{C}{2} as well as
 the [\ion{Co}{2}] 10.52~\microns line. The N1def model has a strong
 [\ion{Co}{3}] 11.88~\microns feature. Interestingly, the N1def model shows
 a feature around 12.8~\microns that appears almost purely due to
  the [\ion{Ne}{2}] 12.81 line with little evidence for blending from the [\ion{Fe}{3}] 12.84 line.
 The model [\ion{Ne}{2}] feature is clearly shifted to the blue
 compared to the
 observed feature, strengthening 
 the case for the  identification of that
 feature as due to [\ion{Fe}{3}] and not [\ion{Ne}{2}].

\section{Discussion}
\label{sec:disc}

\subsection{Atomic Lines}

\begin{figure*}[ht]
    \centering
    \gridline{\fig{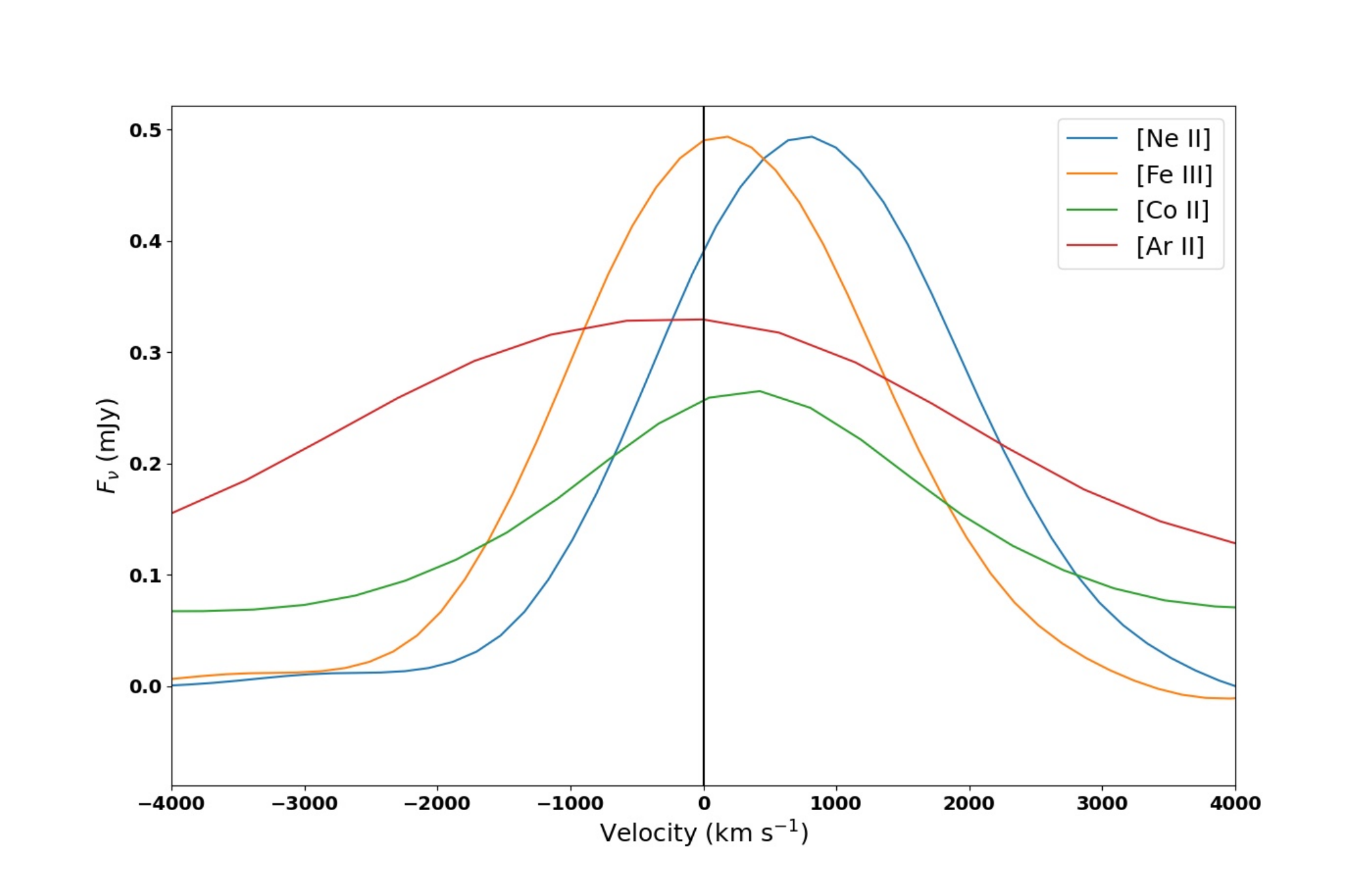}{0.48\textwidth}{(a)}\fig{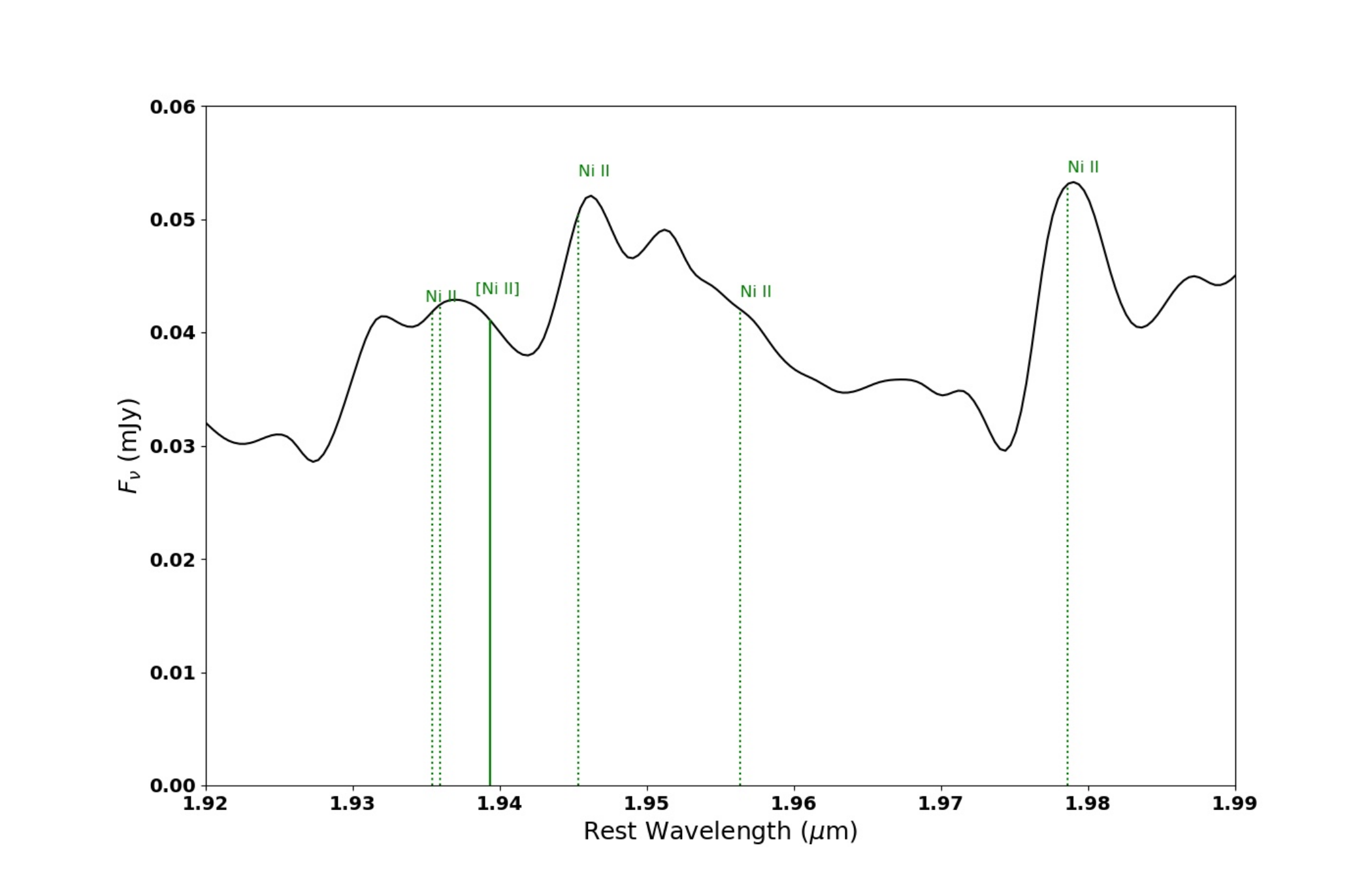}{0.48\textwidth}{(b)}}
    \caption{(a) The velocity space extent of the observed [\ion{Co}{2}]
      10.52, [\ion{Ne}{2}] 
      12.81, and [\ion{Fe}{3}] 12.84 \microns lines. The 12.8 feature
      is the same observed feature, simply shifted to be identified
      with [\ion{Ne}{2}] in one case and with [\ion{Fe}{3}] in the
      other case. (b) A zoomed-in view of the 1.9~\microns region of
      the \vjm spectrum. The [\ion{Ni}{2}] 1.94~\microns line does not
      match the observed features. The strong permitted lines
      ($\log(gf) > 0$) lines shown match some features, in particular
      the observed 1.98~\microns feature.
    \label{fig:co_vs_ne}.}
\end{figure*}

\citet{Kwok:2025} identified permitted lines of iron group elements \ion{Fe}{2}/III, \ion{Co}{2}/III, \ion{Ni}{2}, \ion{Ti}{2},  permitted lines of \ion{O}{1}, \ion{Si}{2}/III,
\ion{Ca}{2}, as well as forbidden [\ion{Fe}{2}], [\ion{Co}{2}], [\ion{Ni}{2}/III], [\ion{Mg}{2}], [\ion{Ne}{2}], [\ion{Ar}{2}], [\ion{Ca}{4}]. 
We find similar IDs, except we don't find neon or magnesium.
Primarily, we  find
lines due to [\ion{Ar}{2}/III], [\ion{Ca}{2}], [\ion{Fe}{2}/III], [\ion{Co}{2}/III], [\ion{Ni}{2}]
and permitted lines from the same ions.
Note that in \autoref{fig:quickid} 
the line near 12.8~\microns is identified as due to [\ion{Fe}{3}]
12.84~\microns  and not the  
[\ion{Ne}{2}] resonance line, since the observed feature is significantly
redward of the [\ion{Ne}{2}] 12.81~\microns
wavelength. \autoref{fig:co_vs_ne}(a) shows the observed spectrum in velocity
space. The velocity extent of the
[\ion{Co}{2}] 10.52 \microns resonance line, the [\ion{Ne}{2}] 12.81 \microns
resonance line (assuming the 12.8~\microns feature is due to [\ion{Ne}{2}]),
and the [\ion{Fe}{3}] 12.84 \microns resonance line (assuming the 12.8~\microns feature is due to [\ion{Fe}{3}]). The
velocity extent of the [\ion{Co}{2}] line is similar to that of the putative
[\ion{Fe}{3}] line, while if the 12.8~\microns feature is due to
[\ion{Ne}{2}] it would have to be significantly  more redshifted from
its rest wavelength. 
\citet{Kwok:2025} identified a feature near 12.8~\microns as due to the
[\ion{Ne}{2}] 12.81~\microns line, which is narrow and centrally
peaked, concluding that since neon is produced in low density regions,
the neon must be thoroughly mixed, therefore the explosion must be due
to a pure deflagration and not a violent merger. 

\subsubsection{Stable Nickel}

Importantly, we also identify the [\ion{Ni}{2}] 6.636 resonance line  and 
\ion{Ni}{2} lines near 1.09~\microns at this
epoch indicating the presence of stable nickel. This is a strong test
for models, since stable nickel implies burning at high densities
where the electron capture rates are large enough to neutronize the
material.

\subsubsection{The 1.94~\microns feature}
In \autoref{fig:quickid} we identify a feature at 1.98~\microns with
\ion{Ni}{2}. A similar feature was identified in \sneia by \citet{Friesen:2014} as
due to the [\ion{Ni}{2}] 1.94~\microns doublet, even though the
feature is significantly redward of the rest wavelength of the
doublet. Recent analysis of a set of NIR spectra of \sneia, 
found evidence for  the [\ion{Ni}{2}] 1.94~\microns doublet
\citep{Kumar:2026}. In their sample, the observed feature tended to be
redshifted, but not by the 0.04~\microns found by
\citet{Friesen:2014}. \autoref{fig:co_vs_ne}(b) presents a zoomed in
view of the \vjm spectrum, focusing on the region around
1.94~\microns. It shows that the [\ion{Ni}{2}] 1.94~\microns doublet
does not well match any of the observed features seen in the
spectrum. Also shown in \autoref{fig:co_vs_ne}(b) are strong ($\log(gf) \ge 0$),
high-excitation \ion{Ni}{2} permitted lines. One of the lines is
reasonably well identified with the observed P-Cygni feature near
1.98~\microns, and a \ion{Ni}{2} 1.9454~\microns line is also
plausibly associated with an observed P-Cygni feature, however the
\ion{Ni}{2} 1.9563~\microns line is not associated with a strong
observed feature and the \ion{Ni}{2} 1.9353 and 1.9359~\microns lines together
with the [\ion{Ni}{2}] 1.94~\microns doublet could be producing
another observed emission feature. These features along with the
[\ion{Ni}{2}] 6.64~\microns resonance line make a case for the presence of stable
nickel in the \vjm ejecta. However, the specific formation of the
lines in the 1.94~\microns region is likely due to blends of nickel and
cobalt lines. For the 1.98~\microns feature, there is a \ion{Co}{2}
line, but it is much weaker, with $\log(gf) \sim -2$, thus it is
likely that the feature is predominantly due to \ion{Ni}{2} with a
small contribution from \ion{Co}{2}.

\subsection{Model Comparison}

Comparison of the models shown in 
\autoref{fig:CO_iso_T2000_wdep_nomol} which include
molecular opacity for both the \wsiax and N1def model with their
counterparts neglecting molecular opacity 
makes it clear that molecular opacity plays an important role in
producing the model spectra.
While the contributions from the fundamental and first overtone of CO
are obvious in both models, both models also produce SiO. For the
N1def model the contributions from CO and SiO seem to account for most
of the molecular emission, while for the \wsiax model, the entire SED
is dominated by molecular features not limited to just the fundamental
and first overtones of CO and SiO. The \wsiax model produces a
significant amount of SiS, whose fundamental is at about 14.2~\microns, but clearly other molecular opacity is
important in the synthetic spectrum.

\subsubsection{SiO and SiS}
\begin{figure*}[ht]
    \centering
    \gridline{\fig{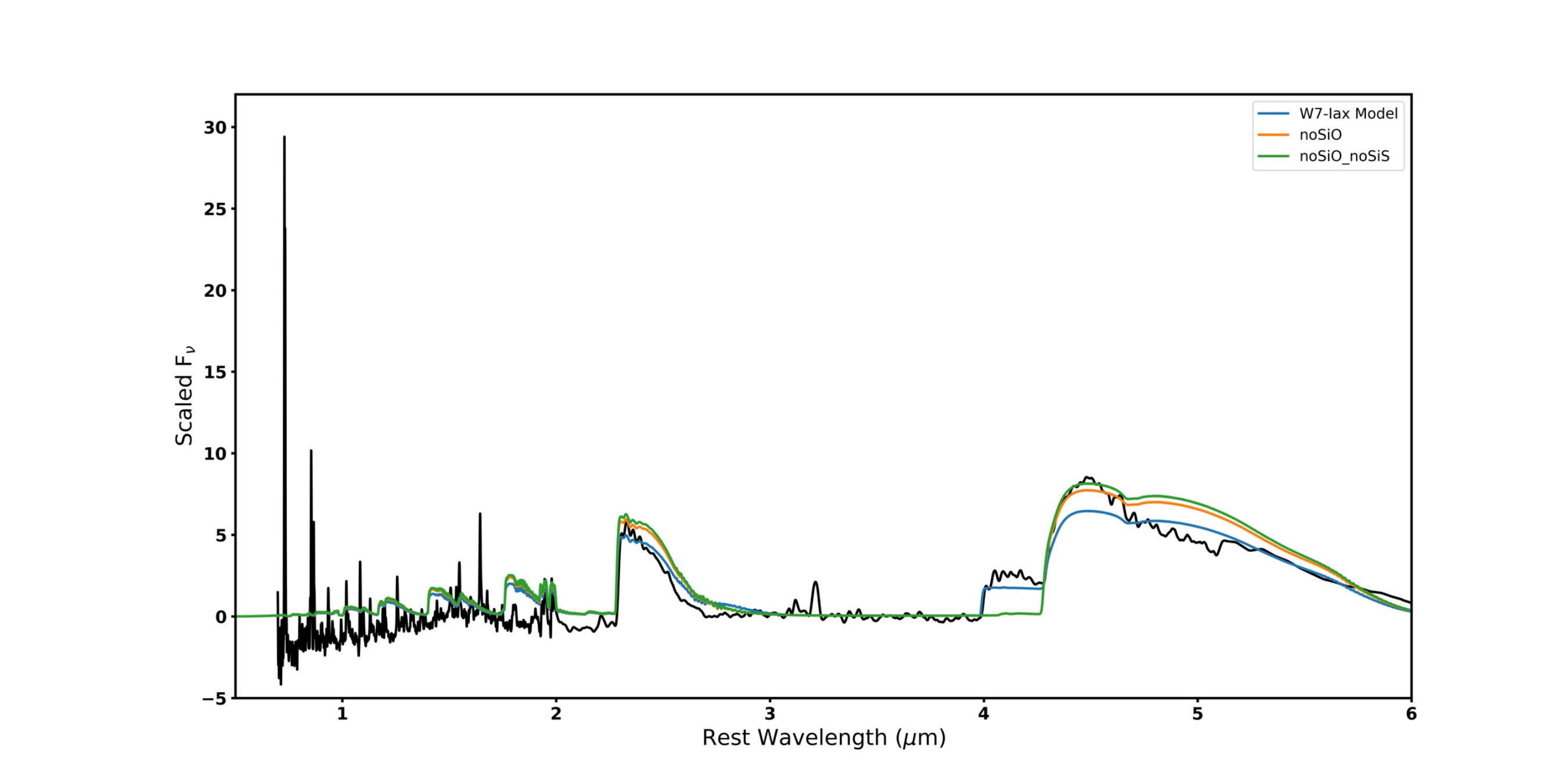}{0.9\textwidth}{(a)}}
    \gridline{\fig{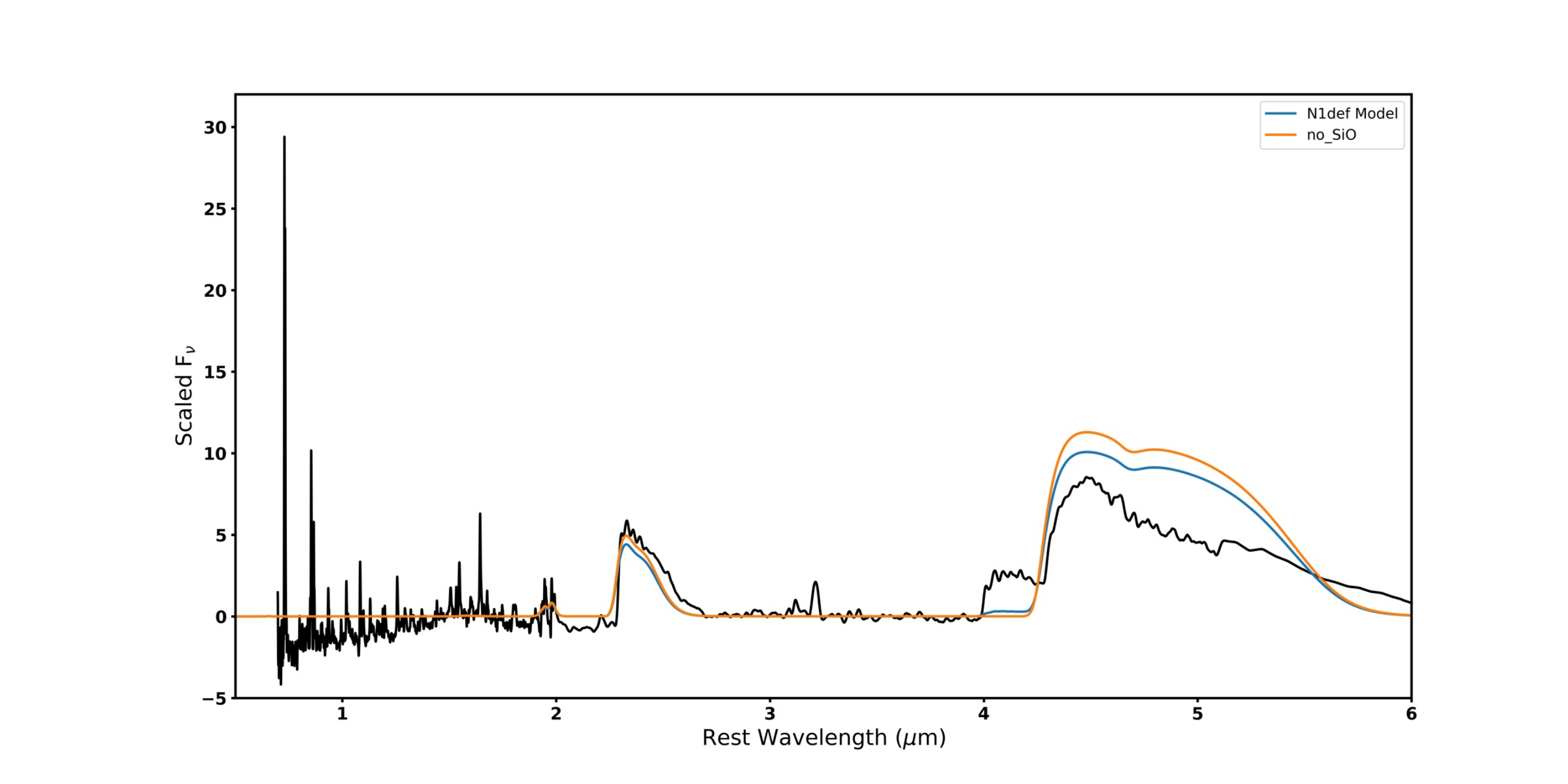}{0.9\textwidth}{(b)}}
\caption{(a) The \wsiax model shown in \autoref{fig:CO_iso_T2000_wdep}(a), compared
  to the same model with the SiO opacity turned off and with both the
  SiO and SiS opacities turned off.
  (b) The N1def model shown in
      \autoref{fig:CO_iso_T2000_wdep}(b), is compared to the same model
      with the SiO opacity turned off.
    \label{fig:N1DEF_iso_T2000_wdep_noSiO}      }
\end{figure*}

\autoref{fig:N1DEF_iso_T2000_wdep_noSiO}(a) shows a comparison of the
synthetic spectrum, 
for the \wsiax model, the same spectrum with SiO opacity turned off,
and the same spectrum with both 
SiO and SiS opacity turned off. Interestingly, the flat band
seen in the \vjm spectrum from $\sim 4-4.25$~\microns is produced by
the SiO first overtone, so we can confidently identify the presence of
SiO as well as CO in \vjm. A feature in the same wavelength range has
been attributed to [\ion{Ca}{5}] by 
\citet{Kwok:2025b} in the normal \sneia 2022gy and 2022aaiq, but here
we are sure that it is due to the SiO first overtone. Since the first overtone of SiO is present,
the fundamental band of SiO also plays a role, but it is more
difficult to match the features produced in the synthetic spectrum
with that of the observed.
The SiS creates a strong
feature in the $\sim 6.5-7.5$~\microns range (from the first overtone), but since the model
does not produce the strong forbidden lines in this region  it is hard to
definitively identify or rule out the presence of SiS.  
\autoref{fig:N1DEF_iso_T2000_wdep_noSiO}(b) shows the N1def model 
spectrum with the opacity of SiO turned off. 
The stronger
SiO first overtone in the \wsiax model compared to that of N1def,
more closely fits the observed \vjm spectral feature, but since the
strong forbidden lines are missing for both models, it is premature
for 
definitive statements about which model is favored.
Nevertheless, it seems reasonable to conclude that mass of CO and SiO
produced by the \wsiax model approximately correspond to the mass
produced in the \vjm ejecta, in particular that the SiO mass should be
closer to $10^{-3}$~\Msun than to $10^{-4}$~\Msun.

\subsection{Previous CO identifications in Thermonuclear SNe}

\citet{Hsiao:2020_lsq14fmg} saw a steep drop in the light curve of the
2003fg-like LSQ2014fmg around a month post maximum light that they
attributed to the formation of CO. At the same epoch, there is no
evidence for a steep drop in the light curve of \vjm. However, \vjm
went behind the sun about 2 months post maximum, so that drop likely
occurred later, with the onset of CO formation.

\section{Conclusions}

We obtained a spectrum of the dim \sniax \vjm with JWST on day +\phasebmaxrf,
that combined with \jwst imaging from day 196.9 allows the construction
of an SED  with coverage from  0.7--21~\microns. The observed
spectrum shows strong permitted and  forbidden lines of intermediate mass (Ca
and Ar) and iron-group elements (Ni, Co, Fe). While the [\ion{Ne}{2}]
12.81~\microns resonance line may be present, there are no other
obvious neon lines and [\ion{Fe}{3}] fits the feature just as well, in
addition to fitting other observed features, strengthening the 12.84 [\ion{Fe}{3}]
identification.
This makes  the absence of neon an important model discriminator. The
observed SED shows strong emission from both the CO fundamental and
first overtone as well as a pronounced feature that is well fit by the
first overtone of SiO, indicating that \vjm has formed both CO and SiO
by this epoch. The presence of [\ion{Ni}{2}] as well as permitted
\ion{Ni}{2} lines at this late epoch
implies that densities are high enough that electron capture is
effective, a strong constraint on explosion scenarios, in that  the
high densities must occur either in the progenitor or material must be
ejected from the bound remnant (and likely efficiently mixed into the ejecta).

The line widths of the iron group elements are narrower and
therefore are present at lower velocities than, in particular, the
[\ion{Ar}{2}] and [\ion{Ar}{3}] lines, indicating radial
stratification. On the other hand, the [\ion{Ar}{2}] and [\ion{Ar}{3}]
lines are not flat-topped, indicating that they are not confined to a
shell. Furthermore, the [\ion{Ca}{2}] doublet  is resolved, with each
component symmetric around about zero velocity with a velocity width
of $\sim 500$~\kmps, significantly smaller than the $\sim 3000$~\kmps
width of the argon features. This could possibly be due to the
formation of the calcium features in a circumstellar wind.

The simple dust model constructed argues for a small amount of
carbonate dust. While a single epoch cannot determine whether the dust
is newly formed or pre-existing, we have made a case that the dust is pre-existing.

We calculated synthetic spectra of two characteristic explosion
models: a modified version of W7, \wsiax, where the ejecta mass and velocities
have been artificially reduced, but the compositions unaltered and
N1def, a pure deflagration model that leaves behind a bound remnant. In
both cases our synthetic model spectra produced forbidden line
features that are much weaker than seen in the observed spectrum, but
molecular features that agree reasonably well with those observed.
Both models produced significant emission from CO and SiO,
while the \wsiax model also produced SiS. The lack of fidelity of the
model to the observations makes it hard to rule out the presence of
SiS, but it doesn't seem to be needed. If SiS could be definitively
ruled out, the molecular features would seem to favor a fully mixed
model like N1def over a radially stratified model like the \wsiax model,
since in N1def SiO forms at the highest density, inhibiting the
formation of SiS, whereas in the \wsiax model SiS and SiO form
just below the C+O shell. However, the \wsiax model reproduces
the SiO first overtone significantly better than does N1def,
suggesting that the amount of  SiO produced by the \wsiax
model is approximately the amount in the ejecta of \vjm. \wsiax does
also produce significant molecular absorption features blueward of two
microns that are absent in both N1def and the observed spectrum. Thus,
definitive conclusions about whether the ejecta of \vjm are more
uniform like that of N1def or more radially stratified like that of the \wsiax
model are premature.

Future work will focus on obtaining stronger forbidden lines, defining
a criterion to determine the presence or absence of a bound remnant,
the formation of in-situ dust, 
and the presence or absence of neon.

\section{Acknowledgments}

We thank Lindsey Kwok for generously sharing her reduced data with us.
Some of the data presented in this paper were obtained from the
Mikulski Archive for Space Telescopes (MAST) at the Space Telescope
Science Institute. The specific observations analyzed can be accessed
via
\dataset[https://doi.org/10.17909/fs4r-tk12]{https://doi.org/10.17909/fs4r-tk12}. STScI 
is operated by the Association of Universities for Research in
Astronomy, Inc., under NASA contract NAS5–26555. Support to MAST for
these data is provided by the NASA Office of Space Science via grant
NAG5–7584 and by other grants and contracts.
E.B., C.A., J.D., M.S., and  P.H. acknowledge support from NASA grants JWST-GO-02114,
JWST-GO-02122, JWST-GO-04522, JWST-GO-04217, JWST-GO-04436,
JWST-GO-03726, JWST-GO-05057, JWST-GO-05290, JWST-GO-06023,
JWST-GO-06677, JWST-GO-06213, JWST-GO-06583. Support for
programs \#2114, \#2122, \#3726, \#4217, \#4436, \#4522,  \#5057,
\#6023, \#6213, \#6583, and \#6677
were provided by NASA through a grant from the Space Telescope Science
Institute, which is operated by the Association of Universities for Research in
Astronomy, Inc., under NASA contract NAS 5-03127.
E.B.,  C.A., and J.D. acknowledge support from HST-AR-17555, Support for
Program number 17555 was provided through a grant from the STScI under NASA
contract NAS5-26555.
E.B. thanks the Yukawa Institute for Theoretical Physics at Kyoto
University. Discussions during the YITP long-term workshop
YITP-T-26-02  on ``Multi-Messenger Astrophysics in the Dynamic Universe'' were useful to complete this work.
J.L. acknowledges support from NSF grant AAG-2206523.
L.G. acknowledges financial support from the Spanish Ministerio de
Ciencia e Innovaci\'on (MCIN), the Agencia Estatal de Investigaci\'on (AEI)
10.13039/501100011033, and the European Social Fund (ESF) "Investing in your
future" under the 2019 Ram\'on y Cajal program RYC2019-027683-I and the
PID2020-115253GA-I00 HOSTFLOWS project, from Centro Superior de Investigaciones
Cient\'ificas (CSIC) under the PIE project 20215AT016, and the program Unidad
de Excelencia Mar\'ia de Maeztu CEX2020-001058-M, and from the Departament de
Recerca i Universitats de la Generalitat de Catalunya through the
2021-SGR-01270 grant.
M.D. Stritzinger is funded by the Independent Research Fund Denmark (IRFD,
grant number 10.46540/2032-00022B).
Some of the calculations presented here were performed at the National Energy
Research Supercomputer Center (NERSC), which is supported by the Office of
Science of the U.S. Department of Energy under Contract
No. DE-AC03-76SF00098.
The authors gratefully acknowledge the computing time
made available to them on the high-performance computers HLRN-IV
at GWDG at the NHR Center NHR\@G\"ottingen and at ZIB at the NHR
Center NHR\@Berlin. These Centers are jointly supported by the Federal
Ministry of Education and Research and the state governments
participating in the NHR (\url{https://www.nhr-verein.de/unsere-partner}). 
We also thank OU
Supercomputing Center for Education \& Research (OSCER) at the University of
Oklahoma (OU).

\clearpage

\bibliography{sn24vjm_DDT}{}

\end{document}